# Unwinding Conditional Noninterference

Chenyi Zhang

*Faculty of Sciences, Technology and Communication, University of Luxembourg*
*6, rue Richard Coudenhove-Kalergi, L-1359 Luxembourg*

*Abstract*—Noninterference provides a control over information flow in a system for ensuring confidentiality and integrity properties. In the literature this notion has been well studied as transitive noninterference and intransitive noninterference. In this paper we define a framework on the notion of conditional noninterference, which allows to specify information flow policies based on the semantics of action channels. Our new policies subsume the policies of both transitive and intransitive noninterference, and support dynamic requirements such as upgrading and downgrading. We also present unwinding relations that are both sound and complete for the new policies.

*Keywords*-Information Flow; Noninterference; Unwinding;

## I. INTRODUCTION

Information flow security policies are concerned with both confidentiality and integrity requirements of a system. The seminal work by Goguen and Meseguer introduces a way of defining information flow security policies by a set of *noninterference* assertions [17]. Each assertion specifies that a given set of actions are *not allowed to interfere with* a security domain. The follow-up works often interpret a noninterference policy as a *relation* over a set of security domains indicating *permitted flow* of information. If a policy relation is transitive, it has a natural correspondence to the classical multilevel security policies of Bell and La-Padula [3], [4]. Therefore, until recently, most work in this area defines a policy on how to allow information to flow among security domains, instead of how to disallow such flow as explored in the original paper.

The transitive noninterference policies are sometimes considered as too strong in many situations, because they require that information flow is totally blocked from one security domain to another at any time. A weakened version of noninterference is to allow a policy relation to be intransitive [32], [31], [42], [38]. This makes it possible to specify a more flexible flow policy. For example, one may define a policy $\leadsto \subseteq \{A, B, C\} \times \{A, B, C\}$ for a system with three security domains, such that domain $A$ is allowed to send information to domain $B$ by $A \leadsto B$ (i.e., $(A, B) \in \leadsto$), and that domain $B$ is allowed to send information to domain $C$ by $B \leadsto C$ (i.e., $(B, C) \in \leadsto$). However, domain $A$ is *not* allowed to directly send information to $C$ if $(A, C)$ is not in the policy relation $\leadsto$. In this case $B$ may be regarded as a channel that controls information flowing from $A$ to $C$, which is not expressible by the original (transitive) noninterference policies [32]. The notions of transitive and intransitive noninterference have been applied in different areas such as operating system verification [19], [26], security protocol verification [1], [15], and programming language analysis [33], [34].

However, it is also in the paper of Goguen and Meseguer [17] that another weakened form called *conditional noninterference* was proposed. Conditional noninterference associates each noninterference assertion with a constraint, in the way of $A \not\leadsto u \; [\![\phi]\!]$, such that the noninterference assertion takes effect (i.e., $A$ becomes invisible to $u$, as for *confidentiality*, or $A$ is not allowed to change $u$, as for *integrity*) whenever the constraint $\phi$ is satisfied. In other words, $A \not\leadsto u$ is conditional to $\phi$. Although this notion is not followed in subsequent works in the information flow literature (to our knowledge), it proposes an insight that it is also viable to place a control before information flow is allowed to happen. Note that intransitive noninterference only specifies how to allow information propagation after an action of intended flow occurs.

In this paper, we present a policy framework for *conditional noninterference* to incorporate both intransitive noninterference [21] and the notion of the same name as presented by Goguen and Meseguer [17]. (We overload this term because we believe it carries the appropriate meaning.) We are going to show that the noninterference assertions with the additional conditions can be used to express not only the channel control policies, but also some other useful security requirements, including a certain class of policies for dynamic control. From the perspective of channel control, our framework turns out more general than intransitive noninterference in different ways.

*Unwinding theorems* [18], [32], [42] are useful techniques to verify noninterference-based properties. Given a set of noninterference constraints, it is possible to define a set of *unwinding relations* for each user (or security domain), so that if the relations satisfy a number of constraints, then it is sufficient to say that a system is secure. Unwinding is a very desirable technique since it reduces verification of noninterference properties into conditions that are easily provable by existing tools, with available examples applying theorem provers PVS [10] and Isabelle/HOL [42]. The unwinding theorem for deterministic state based systems is also complete if the underlying security policy on *interference* (i.e., the induced binary relation on the set of users) is transitive [32]. However, the (weak) unwinding relations in the literature [32], [42] are not necessary conditions for intransitive noninterference even if the system is deterministic. In fact, the weak unwinding relation for intransitive noninterference rather corresponds more or less to a notion that is strictly stronger than the intransitive noninterference properties [38]. In this paper, we define unwinding relations for more general classes of noninterference properties which subsume intransitive noninterference. Nevertheless, we prove that the existence of such unwinding relations are both *sound* and *complete* for a system to be secure with respect to the properties defined in this paper.

The main contributions of this paper are as follows. (1) We apply conditional noninterference to express a variety of security requirements, such as upgrading, downgrading, and channel control. (2) We identify two subclasses of conditional noninterference properties, and for each subclass we design a new unwinding technique which is both sound and complete to the properties in this class in a very general way. (3) As a byproduct, we show that a subclass of our properties can be reduced to safety properties by a doubling construction. The outline of the paper is as follows. In Sect. II we define the system model and rephrase the classical noninterference definition. Sect. III presents conditional noninterference and shows how it can be used to express many interesting security requirements. In Sect. IV we define unwinding techniques to characterize the conditional noninterference properties, and for a particular class of policies, we reduce their verification problems to safety properties. Sect. V discusses related work. Sect. VI concludes the paper and suggests possible future research directions.

## II. Noninterference

We define a state machine model similar to those that one can find in the literature [17], [32]. We assume a (finite) set of *users* (or *security domains*) $U$, a set of actions $\mathcal{A}$, and a function $dom : \mathcal{A} \to U$ that maps each action to a user who performs it. In our model, each action is associated with a unique security domain, since in practice if there is an action that is available to more than one users, we add distinct usernames as subscripts to produce different actions. The tuple $(\mathcal{A}, U, dom)$ is called a *signature*, based on which we write $\mathcal{A}_u$ as the set $\{a \in \mathcal{A} \mid dom(a) = u\}$ for $u \in U$. We write $a, b, a_1, \ldots$ to range over $\mathcal{A}$.

A *machine* for a given signature $(\mathcal{A}, U, dom)$ is a tuple of the form $M = \langle S, s_0, step, obs, O \rangle$ where

- $S$ is a set of states,
- $s_0 \in S$ the initial state,
- $step : S \times \mathcal{A} \to S$ the transition function,
- $obs : U \times S \to O$ the observation function,
- $O$ is a set of outputs.

The function $step$ describes the system transition, such that $step(s, a)$ is the unique next state when action $a$ is applied on state $s$. The function $obs$ gives an observation made in each state by a user. For readability, we 'curry' the function $obs$ by $obs_u$ of type $S \to O$ given $u \in U$. Note that such a machine is always *input enabled* by the definition of function $step$, so that every input action is enabled on every state. Also, a machine is always *deterministic* in the sense that given a state $s$ and sequence of actions $\alpha \in \mathcal{A}^*$, a run of state sequence can be uniquely determined. To denote the final state after the execution of a sequence of actions, define the operation $\bullet : S \times \mathcal{A}^* \to S$, by $s \bullet \epsilon = s$, and $s \bullet (\alpha \cdot a) = step(s \bullet \alpha, a)$ for $s \in S$, $a \in \mathcal{A}$ and $\alpha \in \mathcal{A}^*$. We assume every state in a machine is reachable.

In this model we define observation on states, which is different from the definitions of Rushby [32] where observations are associated with actions. This distinction is not essential for many security notions [39], including noninterference. In literature the state-observed machines have also been used by a number of authors, such as Goguen and Meseguer [17] and Bevier and Young [5]. Our choice on modelling of a machine is arbitrary.

The security policy we are to define assumes a partition on the set of actions. Given a signature $(\mathcal{A}, U, dom)$, define a partition `Part` over $\mathcal{A}$ satisfying the following conditions.

1) For all $P \in$ `Part`, there exists $u \in U$ such that $P \subseteq \mathcal{A}_u$,
2) $\bigcup$ `Part` $= \mathcal{A}$,
3) $P_1 \cap P_2 = \emptyset$ for all distinct $P_1, P_2 \in$ `Part`.

We define a function $part : \mathcal{A} \to$ `Part` that assigns each action a unique partition. Obviously $part$ refines $dom$.

A *noninterference assertion* $T$ is of the form $\langle P \not\rightsquigarrow u \rangle$ for $u \in U$ and $P \in$ `Part`, referring to a security requirement that an action partition $P$ is not allowed to interfere with a user $u$.[1] (With respect to integrity, this assertion could also be interpreted as that actions in $P$ are not allowed to 'touch' $u$, where $u$ may represent a real entity, e.g., a device or a file rather than a user.) In this case we say $T$ controls $P$ and *is associated with* $u$. This definition is intuitively finer than what is presented by Rushby [32] who defines interference (the complement of noninterference) as a relation over the set of users. We choose this structure for noninterference assertions not only because it is seemingly finer and more general, but also it seems more reasonable. When noninterference is used to express complex security conditions, this structure sometimes provides a more reasonable control. For example, a user in charge of downgrading can avoid unnecessary downgrading of information by choosing actions not in the partition of downgrading actions.[2] A noninterference security policy is a set of noninterference assertions, for which we use symbols such as $\Pi, \Pi'$.

---

[1] A similar form can be found in [17] where $u|B \not\rightsquigarrow v$ is used to denote that $u$ is not allowed to interfere with $v$ via the actions in $B$. We simplified the presentation by explicitly defining action partitions to be associated with unique security domains.

Given a security policy $\Pi$ and an action sequence $\alpha \in \mathcal{A}^*$ a function $purge_\Pi : \mathcal{A}^* \times U \to \mathcal{A}^*$ is introduced (as in [17]) to clear away from $\alpha$ the actions that are not allowed to interfere with a security domain $u$, which is inductively defined by $purge_\Pi(\epsilon, u) = \epsilon$, and

$$purge_\Pi(a \cdot \alpha, u) = \begin{cases} purge_\Pi(\alpha, u) & \text{if } part(a) \not\leadsto u \\ a \cdot purge_\Pi(\alpha, u) & \text{otherwise.} \end{cases}$$

A system satisfies noninterference, if for all $u \in U$ and $\alpha \in \mathcal{A}^*$, $obs_u(s_0 \bullet \alpha) = obs_u(s_0 \bullet purge_\Pi(\alpha, u))$. Plainly, this requires that removing all the actions not allowed to interfere with a user is not noticeable by that user, since it gives the same view to that user as the action sequence in which no actions are removed. From the set of noninterference assertions $\Pi$, a relation $\leadsto \subseteq U \times U$ of interference is uniquely determined. Write $u \leadsto v$ if there exists a nonempty set of actions $B \subseteq \mathcal{A}_u$ such that for all noninterference assertions in the form of $\langle P \not\leadsto v \rangle$, we have $P \cap B = \emptyset$. We say that the noninterference policy is transitive if the induced relation $\leadsto$ is transitive on $U$.

Most of the policies studied in literature are transitive. For example, MultiLevel Security of Bell and LaPadula [3] defines a partial order of security domains.[3] Later Denning introduced a lattice structure of security classes to reason about information flow [12]. Noninterference can be used to analyze transitive information flow policies, but it is not necessarily transitive by nature. To be explicit, the relation $\leadsto$ induced by a policy $\Pi$ is *not* inherently transitive according to the definition of function $purge$. We sketch it in the following example.

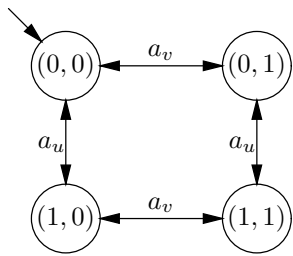

Figure 1. The machine of example 1, where a state $s$ is labelled $(obs_v(s), obs_w(s))$, and we omit the self-transitions by $a_w$.

---

[2]As in the case of a channel control policy [21], [32] where $u \leadsto v$ and $v \leadsto w$, it seems more realistic to let only a subset of $\mathcal{A}_v$ act as a channel passing information from $u$ to $w$. This may also be partially used to defend criticisms against the purge-based channel control policies such as those from Roscoe and Goldsmith [31].

[3]More precisely, it is defined as a combination of a totally ordered set of security labels $\mathcal{L}$ such as top secret (TS), secret (S), confidential (C), unclassified (U), where $TS > S > C > U$ and a set of categories $\mathcal{C}$, such as Navy, Army and Air Force, which are pairwise incomparable, so that $(l_1, c_1) \leq (l_2, c_2)$ with $l_1, l_2 \in \mathcal{L}$ and $c_1, c_2 \in \mathcal{C}$ iff $l_1 \leq l_2$ and $c_1 = c_2$.

**Example** *1:* Let $U = \{u, v, w\}$, and a flow policy satisfying $u \leadsto v$ and $v \leadsto w$, i.e., the set of noninterference assertions is $\Pi = \{\langle \mathcal{A}_v \not\leadsto u \rangle, \langle \mathcal{A}_w \not\leadsto v \rangle, \langle \mathcal{A}_w \not\leadsto u \rangle, \langle \mathcal{A}_u \not\leadsto w \rangle\}$, where $\mathcal{A}_u = \{a_u\}$, $\mathcal{A}_v = \{a_v\}$ and $\mathcal{A}_w = \{a_w\}$. Let $S = \{0, 1\} \times \{0, 1\}$ with $s_0 = (0, 0)$, $obs_u((x, y)) = \emptyset$, $obs_v((x, y)) = x$ and $obs_w((x, y)) = y$ for all $x, y \in \{0, 1\}$. The transition function is defined as $step(s, a_w) = s$ for all $s$, $step((x, y), a_u) = (x \otimes 1, y)$ and $step((x, y), a_v) = (x, y \otimes 1)$ for all $x, y \in \{0, 1\}$, where $\otimes$ denotes 'exclusive or'. The machine is depicted in Fig. 1.

One may observe that $u$ determines $v$'s observation and $v$ determines $w$'s observation in every state. Every noninterference assertion can be verified via the *purge* function. For instance for the assertion $\langle \mathcal{A}_u \not\leadsto w \rangle$, we have $obs_w(s_0 \cdot \alpha) = obs_w(s_0 \cdot purge_\Pi(\alpha, w))$ for all $\alpha \in \mathcal{A}^*$. Note that the relation $\leadsto$ is not transitive, since $(u, w) \notin \leadsto$, although we have $u \leadsto v$ and $v \leadsto w$. □

We claim that Goguen and Meseguer's noninterference policy is not necessarily transitive, and moreover, it can be used to encode security properties *stronger* than those that are known as intransitive noninterference or channel control policies [21], [32]. In the above example $v$ can pass information from $u$ to $w$ only after he indeed *receives* and *knows* the information. Furthermore, $v$ is allowed to intentionally *block* information from $u$ to $w$, although information is free to flow from $u$ to $v$ and from $v$ to $w$. This example provides a view on the notion of noninterference of Goguen and Meseguer that it also gives a channel-like control which works differently from that of intransitive noninterference. Note that in intransitive noninterference policies, it is possible that a channel is allowed to forward information without knowing what is being forwarded [38].

## III. CONDITIONAL NONINTERFERENCE

Conditional noninterference was introduced to support dynamic policies [17], where the conditions were predicates on a sequences of actions *before* reaching a state. In this section the notion is extended also to the other direction (similar to intransitive noninterference), so that conditional noninterference decides whether an action is allowed to interfere with a user given a path of actions leading to the current system state as well as the possible future actions to be performed.

We define *conditional noninterference assertion* to be of the form $\langle P \not\leadsto u \ [\![ \phi ]\!] \rangle$, where the condition $\phi$ is a function of type $\mathcal{A}^* \times \mathcal{A} \times \mathcal{A}^* \to \{true, false\}$. Given a sequence of actions $\alpha \in \mathcal{A}^*$, a single action $a \in \mathcal{A}$, and another sequence of actions $\alpha' \in \mathcal{A}^*$ to be executed in the future, $\phi(\alpha, a, \alpha')$ answers whether the current action $a$ is allowed to interfere with user $u$, i.e., whether it needs to be 'purged'. The sequence $\alpha$ can be understood as the pre-conditional part of the whole sequence $\alpha \cdot a \cdot \alpha'$ for $\phi$, so that a decision is made based on history. The sequence $\alpha'$ represents the actions yet to be performed. This part enables us to define a policy that permits information flow only after it is checked by other users, which has been already explored in the form of intransitive noninterference or channel control policies [21], [32]. We regard $\alpha'$ as the post-conditional part of $\alpha \cdot a \cdot \alpha'$ for $\phi$ on action $a$. If $\phi$ is always evaluated true in assertion $T = \langle P \not\hookrightarrow u \, [\![\phi]\!]\rangle$, then $T$ is a strict assertion, and it is equivalent to what is defined in the previous section. To this point we revise the notion of *security policy* to be a set of conditional noninterference assertions. We have the following definition for the new *purge* function.

**Definition 1:** Given a policy $\Pi$, the function $purge_\Pi : \mathcal{A}^* \times U \rightarrow \mathcal{A}^*$ is defined as for all $\alpha \in \mathcal{A}^*$ in the form of $a_1 a_2 \ldots a_n$, $purge_\Pi(\alpha, u) = a'_1 a'_2 \ldots a'_n$, such that for every $i \in \{1, \ldots, n\}$,

$$a'_i = \begin{cases} \epsilon & \text{if there exists } \langle part(a_i) \not\hookrightarrow u \, [\![\phi]\!]\rangle \in \Pi \\ & \text{and } \phi(\alpha_{-i}, a_i, \alpha_{+i}), \\ a_i & \text{otherwise}. \end{cases}$$

where $\alpha_{-i} = a_1 \ldots a_{i-1}$, $\alpha_{+i} = a_{i+1} \ldots a_n$, and $\epsilon$ denotes the empty sequence of actions.

A system is secure with respect to a policy $\Pi$, if for all $u \in U$ and $\alpha \in \mathcal{A}^*$, $obs_u(s_0 \bullet \alpha) = obs_u(s_0 \bullet purge_\Pi(\alpha, u))$. This requires that every user $u \in U$ is unable to distinguish trace $\alpha$ and $purge_\Pi(\alpha, u)$ by his observations.

In the rest of the section, we restrict our attention to two subclasses of conditional noninterference assertions. For each class of assertions, we define its corresponding *purge* functions. We will also show how these assertions can be applied to express a few existing policies of interest.

### A. Pre- and Post-Conditional Assertions

We define two subclasses of conditional assertions. A pre-conditional assertion provides a control when a decision on permitted information flow needs to be made ahead of time. For example, in a system with discretionary access control, if a user wishes to receive information from a different user, he may simply create a file which he can read, and delegate the 'write' access of this file to that particular user. He may also revoke this access in the future. A post-conditional assertion controls flow of information after an action with intended flow is performed. An example for this policy is that a secret message must be followed by an encrypting action before it is allowed to be sent out. Note that in many circumstances, such decisions on permissions of information passage can only be made by a super-user or an administrator.

We start with a simple language $\Phi^-$ for expressing the pre-conditional and post-conditional assertions as shown in Fig. 2. The superscripts '*pre*' and '*post*' denote whether a constraint is defined in a pre-conditional or post-conditional assertion, and the arrows '$\nearrow$' and '$\searrow$' denote upgrading channels and downgrading channels, respectively. A post-conditional assertion only asserts a condition under which an already-taken action is allowed to produce effect. For example, the assertion $\langle P \not\hookrightarrow u \, [\![[P_1 P_2]_\searrow^{pre}]\!]\rangle$ disallows partition $P$ to interfere with $u$ unless it is immediately

$$\phi_{pre} := [\overleftarrow{C}_1 \cup \overleftarrow{C}_2 \cup \cdots \cup \overleftarrow{C}_n]_\nearrow^{pre} \mid [\overleftarrow{C}_1 \cup \overleftarrow{C}_2 \cup \cdots \cup \overleftarrow{C}_n]_\searrow^{pre}$$
$$\phi_{post} := [\overrightarrow{C}_1 \cup \overrightarrow{C}_2 \cup \cdots \cup \overrightarrow{C}_n]_\rightarrow^{post}$$
$$\overleftarrow{C} := C \mid C \Diamond \mid \overleftarrow{C} C \mid \overleftarrow{C} C \Diamond \qquad C := P \mid P \cup C$$
$$\overrightarrow{C} := C \mid \Diamond C \mid C \overrightarrow{C} \mid \Diamond C \overrightarrow{C} \qquad P := P_1 \mid P_2 \mid \ldots \mid P_n$$

where $P_i \in \texttt{Part}$ for $1 \leq i \leq n$ for some $n$

Figure 2. syntax of the constraints in $\Phi^-$

preceded by an action in $P_1$ followed by an action in $P_2$, and the assertion $\langle P \not\hookrightarrow u \, [\![[\Diamond(P_1 \cup P_2)]_\rightarrow^{post}]\!]\rangle$ allows actions from $P$ to be detectable by $u$ only if somewhere in the future an action in $P_1 \cup P_2$ is performed. In this case, the symbol '$\Diamond$' resembles its usage in temporal logics, in the sense that the actions in the next partition (or union of partitions) are not necessarily to happen immediately after, but *within a finite distance* in the action sequence. We define the post-conditional assertions in the way of controlled release of information, and such release is regarded as *irreversible*.[4]

For the semantics of $\Phi^-$, every channel inside an assertion is interpreted as a regular expression. Let $[\![.]\!]$ be a function from $\Phi^-$ to regular expressions. For a channel constraint $\overleftarrow{C}_i = W_1 W_2 \ldots W_n$ (or $\overrightarrow{C}_i$ for the post-conditional case) where $W_i \in \{\Diamond\} \cup \mathcal{P}(\mathcal{A})$, define $[\![\overleftarrow{C}_i]\!]$ as the regular language represented by $W'_1 W'_2 \ldots W'_n$ where $W'_i = \mathcal{A}^*$ if $W_i = \Diamond$ and $W'_i = W_i$ otherwise. Given $\phi = \overleftarrow{C}_1 \cup \overleftarrow{C}_2 \cup \ldots \overleftarrow{C}_n$, we have $[\![\phi]\!] = [\![\overleftarrow{C}_1]\!] \cup [\![\overleftarrow{C}_2]\!] \cup \ldots [\![\overleftarrow{C}_n]\!]$, i.e., the union of the languages of all the *channels*. The semantics for post-conditional constraints are defined in a similar way. Given an assertion $\langle P \not\hookrightarrow u \, [\![\phi]\!]\rangle$, $\alpha, \alpha' \in \mathcal{A}^*$ and $a \in \mathcal{A}$,

- if $\phi$ is in the form of $[\phi']_\nearrow^{pre}$, then $\phi(\alpha, a, \alpha') = true$ iff $\alpha \in \mathcal{A}^*[\![\phi']\!]$,
- if $\phi$ is in the form of $[\phi']_\searrow^{pre}$, then $\phi(\alpha, a, \alpha') = false$ iff $\alpha \in \mathcal{A}^*[\![\phi']\!]$,
- if $\phi$ is in the form of $[\phi']_\rightarrow^{post}$, then $\phi(\alpha, a, \alpha') = false$ iff $\alpha' \in [\![\phi']\!]\mathcal{A}^*$.

Note that the formal interpretation over the upgrading channels and downgrading channels are different. For an upgrading assertion $\langle P \not\hookrightarrow u \, [\![[\phi]_\nearrow^{pre}]\!]\rangle$, if a pre-conditional sequence $\alpha$ matches the pattern, i.e., $\alpha \in \mathcal{A}^*[\![\phi]\!]$, the following action (if in $P$) must be purged. However in the case of downgrading that action must not be purged. The post-conditional assertions only act as downgrading channels.

The usage of the terms 'upgrading' and 'downgrading' are intuitive for both confidentiality and integrity specifications. An upgrading assertion $\langle P \not\hookrightarrow u \, [\![[\phi]_\nearrow^{pre}]\!]\rangle$ allows actions in $P$ to interfere with $u$ (for confidentiality) or $u$ is changeable by $P$ (for integrity) as default, until a pattern in $[\![\phi]\!]$ occurs, after which the policy becomes more strict. An interpretation for downgrading assertions could be made in a similar way. Plainly, every conditional assertion is weaker than its corresponding strict assertion that is generated by removing its conditional part.

---
[4]On the other hand, pre-conditional assertions are allowed to revoke a "permission" to cause flow as long as the actions under control are not yet performed.

*B. Examples*

We sketch two examples to show that conditional policies can be used to express several useful security requirements related to information flow.

**Example** *2:* (book-keeping) We present a simple example of *well-formed* transactions to ensure data integrity by Clark and Wilson [9]. Assume there is a company with a number of employees. A shared data-base $\mathbb{B}$ is in the company's IntraNet from which every user is allowed to retrieve information. A user can modify $\mathbb{B}$, but this is only allowed immediately after he has registered (or authenticated) himself into the system. This is a basic integrity requirement.

Database $\mathbb{B}$ is modelled as a user with no actions, and its observation on the system is just its contents. For a user $E$, his action set $\mathcal{A}_E$ can be partitioned into the set of reading operations $\mathcal{A}_E^r$, the set of writing operations $\mathcal{A}_E^w$ and the book-keeping action $\{a_E^{bk}\}$. The information flow constraints with respect to the security requirement thus can be stated as follows for each user $E$.

(1) $E$'s reading actions are not allowed to change $\mathbb{B}$, which is the assertion

$$\langle \mathcal{A}_E^r \not\to \mathbb{B} \rangle$$

(2) $E$'s writing actions are allowed to modify $\mathbb{B}$ only if that action occurs immediately after a book-keeping action. An assertion for this rule is

$$\langle \mathcal{A}_E^w \not\to \mathbb{B} \; [\![\{a_E^{bk}\}]_{\searrow}^{pre}]\!] \rangle$$

(3) Finally, the action $a_E^{bk}$ also needs to be constrained. If it is not immediately followed by a write operation, it should not affect any part of the database. So we have

$$\langle \{a_E^{bk}\} \not\to \mathbb{B} \; [\![ [\mathcal{A}_E^w]_{\to}^{post}]\!] \rangle$$

$\square$

The above example illustrates how actions need to be bundled together in order to become a *well-formed* transaction. The book-keeping operation serves as a downgrading action on the integrity level of $\mathbb{B}$, after which the employee $E$ is allowed to modify $\mathbb{B}$. The next example presents an upgrading policy.

**Example** *3:* (conflict of interest) In a small town two sales companies $u$ and $v$, which compete with each other, are seeking helps on their business strategies. There is only one consulting company available in that town. If both $u$ and $v$ connect themselves to the consulting company, it raises the requirement that for each individual consultant $c$, once he contacts one company of $u$ and $v$, he will not be allowed to consult the other, so that he cannot play two-sides. This requirement resembles the Chinese Wall security policy [7].[5] We regard both $u$ and $v$ as users with action sets $\mathcal{A}_u$ and $\mathcal{A}_v$. For each consultant $c$, we assume the set of actions he can do is fixed as $\mathcal{A}_c$, which can further be split into disjoint sets $\mathcal{A}_c^u$ and $\mathcal{A}_c^v$ which are supposed to be used to exchange messages with $u$ and $v$, respectively.

(1) Initially, it is required that the companies $u$ and $v$ are not allowed to leak information to each other, which can be sketched as

$$\langle \mathcal{A}_u \not\to v \rangle \text{ and } \langle \mathcal{A}_v \not\to u \rangle.$$

(2) The actions for $c$ to communicate with $u$ are not supposed to have any effect on $v$, so that $v$'s view over the system should not be changed by actions in $\mathcal{A}_c^u$. Similarly, $\mathcal{A}_c^v$ is not allowed to alter $u$'s view. Therefore we have the following assertions.

$$\langle \mathcal{A}_c^u \not\to v \rangle \text{ and } \langle \mathcal{A}_c^v \not\to u \rangle.$$

(3) Once $c$ starts consulting $u$ (or tries to access $u$), he should be immediately disallowed to communicate with $v$. This is defined over the action partition $\mathcal{A}_c^u$ to company $v$. For the effect from partition $\mathcal{A}_c^v$ on company $u$, we define the same assertion.

$$\langle \mathcal{A}_c^u \not\to u \; [\![ [\mathcal{A}_c^v \Diamond]_{\nearrow}^{pre}]\!] \rangle \text{ and } \langle \mathcal{A}_c^v \not\to v \; [\![ [\mathcal{A}_c^u \Diamond]_{\nearrow}^{pre}]\!] \rangle$$

(4) However, it is also possible that $c$ listens to $u$ before he starts to communicate with $v$, so that he can pass information from $u$ to $v$ in an undesired way. Therefore we disallow actions by $u$ to reveal information to $c$ before $c$ shows his intention to consult $u$. This can be sketched by the following assertions.

$$\langle \mathcal{A}_u \not\to c \; [\![ [\mathcal{A}_c^u \Diamond]_{\searrow}^{pre}]\!] \rangle \text{ and } \langle \mathcal{A}_v \not\to c \; [\![ [\mathcal{A}_c^v \Diamond]_{\searrow}^{pre}]\!] \rangle$$

In this example the actions in $\mathcal{A}_c^u$ upgrade the information flow policy on $\mathcal{A}_c^v$ to $v$, i.e., once an action in $\mathcal{A}_c^u$ is performed, the policy becomes more strict on the actions in $\mathcal{A}_c^v$, and vice versa. A reasonable consequence of this policy is that once a consultant tries to communicate with both companies, he will be forbidden to consult both companies thereafter. $\square$

---

[5]A Chinese Wall policy is concerned with the information flow among all the consultants and consulting companies. It has two basic rules: (1) Each consultant is allowed to access at most one company's files in each conflict of interest class, which is known as *simple security* property. (2) Each consultant can write to a company's files only if he has never accessed any other company's file, which is known as $\star$-property. Here we focus on how to prevent information flow between the companies with respect to a particular consultant. We do not prevent an individual consultant from reading one company's file after he has read the other's, as long as this action does not cause information flow between the two companies, in which sense our policy is weaker than the Chinese Wall policy.

## C. More on Pre-Conditional Assertions

The conditional noninterference assertions based on the constraints defined by $\Phi^-$ in Fig. 2 are easy to understand and use, but it might not be general enough to catch more complicated security requirements. For example, it is not possible to have an assertion by $\Phi^-$ to allow an action to act as both downgrading and upgrading in the way of a power switch. In this section, for pre-conditional assertions, we define a more general policy language to achieve better expressiveness. The policy language $\Phi$ is defined as regular expressions on Part.

$$\phi := \emptyset \mid P \mid \phi \cup \phi \mid \phi \cdot \phi \mid \phi^*$$

where $P \in \texttt{Part}$.

We use $\mathcal{A} \setminus P$ to denote $\bigcup \{P' \in \texttt{Part} \mid P' \neq P\}$. A pre-conditional noninterference assertion is thus in the form of $\langle P \not\rightarrow u \; [\![\phi^{pre}]\!] \rangle$, where $P \in \texttt{Part}$, $u \in U$ and $\phi \in \Phi$. The function $\phi^{pre} : \mathcal{A}^* \times \mathcal{A} \times \mathcal{A}^* \rightarrow \{true, false\}$ is defined as $\phi^{pre}(\alpha, a, \alpha') = true$ iff $\alpha \in L(\phi)$. When it is applied to purge an action sequence, the constraint $\phi^{pre}$ removes every action $a$ in partition $P$ from $\alpha \cdot a \cdot \alpha'$, whenever $\alpha$ is in the regular language $L(\phi)$ expressed by $\phi$ in the pre-conditional assertion $\langle P \not\rightarrow u \; [\![\phi^{pre}]\!] \rangle$. In particular, the constraint $\emptyset$ does not purge any actions, and $\mathcal{A}^*$ purges everything, if they appear within an assertion.

Given a user $u \in U$, we define a partial order relation $\leq_u$ on the set of conditional assertions associated with $u$. Say an action sequence $a_1 a_2 \ldots a_n$ is *contained* in another sequence $\alpha$ if there exists $\alpha_0, \alpha_1, \ldots \alpha_n \in \mathcal{A}^*$ such that $\alpha_0 \cdot a_1 \cdot \alpha_1 \cdot a_2 \cdot \alpha_2 \ldots a_n \cdot \alpha_n = \alpha$. Let $T_1$ and $T_2$ be two assertions associated with $u$, $T_1 \leq_u T_2$ if $purge_{\{T_2\}}(\alpha, u)$ is contained in $purge_{\{T_1\}}(\alpha, u)$ for all $\alpha \in \mathcal{A}^*$. Intuitively, this means assertion $T_2$ is stronger than assertion $T_1$, i.e., the language accepted by the constraint in $T_2$ is a superset of the language accepted by the constraint in $T_1$.

**Lemma 1:** For the general pre-conditional assertions,

$\langle P \not\rightarrow u \; [\![\phi_1^{pre}]\!] \rangle \leq_u \langle P \not\rightarrow u \; [\![\phi_2^{pre}]\!] \rangle$ implies $L(\phi_1) \subseteq L(\phi_2)$.

This further induces an ordering on the set of policies, such that given two policies $\Pi_1$ and $\Pi_2$, $\Pi_1 \leq_u \Pi_2$ if for all $u \in U$, and $T_1 \in \Pi_1$, there exists $T_2 \in \Pi_2$ such that $T_1 \leq_u T_2$.

**Proposition 1:** For every pre-conditional assertion $T_1 = \langle P \not\rightarrow u \; [\![\phi^{pre}]\!] \rangle$ with $\phi \in \Phi^-$, there exists a constraint $\psi \in \Phi$, such that the assertion $T_2 = \langle P \not\rightarrow u \; [\![\psi^{pre}]\!] \rangle$ satisfies $T_1 \leq_u T_2$ and $T_2 \leq_u T_1$.

*Proof:* Trivial, since every pre-conditional constraint in $\Phi^-$ expresses a regular expression. ∎

Note this implies that the distinction between downgrading and upgrading assertions in $\Phi^-$ no longer exists in $\Phi$. Since the regular language is closed under complementation,[6] if $R \in \Phi$ expresses a downgrading channel $[\phi]_\searrow^{pre}$, there always exists another expression $R^- \in \Phi$ expressing $[\phi]_\nearrow^{pre}$, such that $R \cap R^- = \emptyset$ and $R \cup R^- = \mathcal{A}^*$.

[6]The author is not sure if it makes sense to have a more general policy language which might not have such good closure properties, e.g., CFL.

The other direction of Prop. 1 does not hold. Following the claim we made at the beginning of the section, the assertion $\langle P \not\rightarrow u \; [\![((\mathcal{A} \setminus Q)^*(Q(\mathcal{A} \setminus Q)^*Q(\mathcal{A} \setminus Q)^*)^*)^{pre}]\!] \rangle$ allows the actions in $Q$ to act as a switch. Even occurrences of actions in $Q$ disallows $P$ to interfere with $u$, while an odd number of actions in $Q$ allows $P$. This is an assertion that expresses a policy mixed with upgrading and downgrading, which is not expressible by $\Phi^-$.

## D. Avoiding Inconsistencies

*Assertion conflict* happens when two assertions associated with the same user and controlling the same partition disagree on whether an action needs to be purged. To resolve this problem, we may take a more secure choice (as stated in Def. 1) by insisting that an action needs to be purged from a sequence if there exists an assertion that returns $true$. Nevertheless this may cause a policy to be potentially stronger than what is expected by a (careless) policy specifier. Formally, two assertions $T_1 = \langle P \not\rightarrow u \; [\![\phi_1]\!] \rangle$ and $T_2 = \langle P \not\rightarrow u \; [\![\phi_2]\!] \rangle$ are in conflict in a policy, if there exists $\alpha, \alpha' \in \mathcal{A}^*$ and $a \in P$, such that $\phi_1(\alpha, a, \alpha') \neq \phi_2(\alpha, a, \alpha')$. Say a policy is *simple*, if for every $P \in \texttt{Part}$ and $u \in U$, there is at most one assertion that controls $P$ and is associated with $u$. In this paper we only discuss simple policies.

Nevertheless, two conditional assertions may conflict each other according to our intuition of permitted information flow even in a simple policy. For example, let a post-conditional assertion $T_1 = \langle P_1 \not\rightarrow u \; [\![\langle \Diamond P_2 \rangle_\rightarrow^{post}]\!] \rangle$ be an assertion that allows $P_1$ to interfere with $u$ only via a channel provided by $P_2$. This assertion is intuitively conflicting the assertion $T_2 = \langle P_2 \not\rightarrow u \rangle$ which disallows $P_2$ to interfere with $u$ in all circumstances. Since $P_2$ is allowed to control information from $P_1$ to $u$ in $T_1$, the information passed from $P_1$ to $u$ carries a 'permission' from $P_2$, which seems undesirable. We propose the following conditions to monitor this type of inconsistencies from a policy.

**Definition 2:** Given a signature $(\mathcal{A}, U, dom)$ and a partition Part,
- a policy $\Pi$ is *left-consistent* if for all $u \in U$ and for all $\alpha, \alpha' \in \mathcal{A}^*$, $purge_\Pi(purge_\Pi(\alpha, u) \cdot \alpha', u) = purge_\Pi(\alpha \cdot \alpha', u)$,
- a policy $\Pi$ is *right-consistent* if for all $u \in U$ and for all $\alpha, \alpha' \in \mathcal{A}^*$, $purge_\Pi(\alpha \cdot purge_\Pi(\alpha', u), u) = purge_\Pi(\alpha \cdot \alpha', u)$.

Intuitively, suppose the effect of action $a$ depends on the existence of action $b$, then the conditions that determine the effect of $b$ should be consistent with the conditions that determine the effect of $a$. A policy being left-consistent (right-consistent) requires that the existence of every action in a purged sequence is consistent with the existence of every other action appearing to the left (right) of that action in the sequence. Obviously, a simple policy consisting of only strict assertions is both left-consistent and right-consistent.

### E. Encoding Intransitive Noninterference

Intransitive noninterference [21], [32] defines an information flow policy as a (reflexive) binary relation $\rightsquigarrow$ over the set of security domains $U$, where $u \rightsquigarrow v$ indicates that $u$ is allowed to interfere with $v$, and $\rightsquigarrow$ is not necessarily transitive on $U$. The $ipurge$ function of type $\mathcal{A}^* \times U \to \mathcal{A}^*$ can be defined as follows.[7] Given $u \in U$ and $\alpha \in \mathcal{A}^*$ in the form of $a_1 a_2 \ldots a_n$, $ipurge(\alpha, u) = a'_1 a'_2 \ldots a'_n$, such that for every $i \in \{1, \ldots, n\}$,
$a'_i = \begin{cases} a_i & \text{if } a_{i+1} a_{i+2} \ldots a_n \text{ contains an } \textit{interference chain}, \\ \epsilon & \text{otherwise.} \end{cases}$
where an interference chain is a subsequence $b_1 b_2 \ldots b_m$ that is contained in $a_{i+1} a_{i+2} \ldots a_n$, satisfying that $dom(a_i) \rightsquigarrow dom(b_1)$, $dom(b_j) \rightsquigarrow dom(b_{j+1})$ for all $1 \leq j \leq m-1$, and $dom(b_m) \rightsquigarrow u$. A system is secure with respect to intransitive noninterference (of policy $\rightsquigarrow$), if for all $u \in U$ and $\alpha \in \mathcal{A}^*$, we have $obs_u(s, \alpha) = obs_u(s, ipurge(\alpha, u))$.

We show that the conditional noninterference policies subsume the intransitive noninterference policies by using only post-conditional assertions. Given a signature $(\mathcal{A}, U, dom)$ and an intransitive noninterference policy $\rightsquigarrow \subseteq U \times U$, we construct a policy $\Pi(\rightsquigarrow)$ as follows. First we let $\texttt{Part} = \{\mathcal{A}_u \mid u \in U\}$. For every pair of users $u, v \in U$, we construct the set $\texttt{Interf}(u,v) = \{v_1 v_2 \ldots v_n \in U^* \mid u \rightsquigarrow v_1 \rightsquigarrow v_2 \rightsquigarrow \ldots \rightsquigarrow v_n \rightsquigarrow v\}$. In this set we enumerate all possible *interference chains* from user $u$ to user $v$. (This set could be infinite.) Define a condense operator $Cond : 2^{\mathcal{A}^*} \to 2^{\mathcal{A}^*}$ by $Cond(TSet) = \{\alpha \in TSet \mid \forall \alpha' \in TSet : \alpha \text{ contains } \alpha' \Rightarrow \alpha = \alpha'\}$. This operator is to remove all redundant and cyclic chains in a set $\texttt{Interf}(u,v)$, so that the remaining condensed set is minimal. For example if $u \rightsquigarrow v$, then neither the chain $u \rightsquigarrow w \rightsquigarrow v$ nor the chain $u \rightsquigarrow w \rightsquigarrow u \rightsquigarrow v$ will provide any additional information on purging actions in $\mathcal{A}_u$ with respect to user $v$. Moreover, such a condensed set will always be finite provided that $U$ is finite. We define $\Pi(\rightsquigarrow)$ as a set consisting of the following assertions. For all distinct $u, v \in U$,

1) if $\texttt{Interf}(u,v) = \emptyset$, then $\langle \mathcal{A}_u \not\rightsquigarrow v \rangle$ is an assertion in $\Pi(\rightsquigarrow)$,
2) if $\texttt{Interf}(u,v) \neq \emptyset$ and $Cond(\texttt{Interf}(u,v)) \neq \{\epsilon\}$, then $\langle \mathcal{A}_u \not\rightsquigarrow v [\![ [\lambda_1 \cup \lambda_2 \cup \cdots \cup \lambda_n]_{\to}^{post}]\!] \rangle$ is an assertion in $\Pi(\rightsquigarrow)$, where $\{\lambda_1, \lambda_2, \ldots, \lambda_n\} = Cond(\texttt{Interf}(u,v))$.

The correctness of the above construction of $\Pi(\rightsquigarrow)$ is by the following result, with its proof sketch in the appendix.

---

[7]The $ipurge$ function in the original paper of Haigh and Young [21] is defined in a different way, but semantically equivalent to the definition here.

**Proposition** *2:* Given an intransitive noninterference policy $\rightsquigarrow$, for all $u \in U$ and $\alpha \in \mathcal{A}^*$, $ipurge(\alpha, u) = purge_{\Pi(\rightsquigarrow)}(\alpha, u)$.

Next we show that intransitive noninterference policies are always right-consistent.[8] The proof of right-consistency requires the following lemmas, which basically show that the $ipurge$ functions are idempotent and they preserve all the interference chains in the results.

**Lemma** *2:* For all $\alpha \in \mathcal{A}^*$ and $u, v \in U$, the sequence $\alpha$ contains an interference chain from $u$ to $v$ iff $ipurge(\alpha, v)$ contains an interference chain from $u$ to $v$.

**Lemma** *3:* $ipurge(ipurge(\alpha, u), u) = ipurge(\alpha, u)$ for all $\alpha \in \mathcal{A}^*$ and $u \in U$.

**Proposition** *3:* Every intransitive noninterference policy is right-consistent.

*Proof:* Given a policy $\rightsquigarrow$, $u \in U$ and $\alpha, \alpha' \in \mathcal{A}^*$, we show $ipurge(\alpha' \cdot \alpha, u) = ipurge(\alpha' \cdot ipurge(\alpha, u), u)$. We prove by induction on length of $\alpha'$. Base case: $ipurge(\epsilon \cdot ipurge(\alpha, u), u) = ipurge(\epsilon \cdot \alpha, u)$ is by Lem. 3.

Suppose $ipurge(\gamma \cdot ipurge(\alpha, u), u) = ipurge(\gamma \cdot \alpha, u)$ for some $\gamma \in \mathcal{A}^*$, we show the case for $a \cdot \gamma$.

- If $ipurge(a \cdot \gamma \cdot \alpha, u) = a \cdot ipurge(\gamma \cdot \alpha, u)$, then there exists an interference chain in $\gamma \cdot \alpha$ from $dom(a)$ to $u$. W.l.o.g, we write $a_1 a_2 \ldots a_i a_{i+1} \ldots a_n$ to be the chain where $a_1 a_2 \ldots a_i$ is contained in $\gamma$, and $a_{i+1} \ldots a_n$ is contained in $\alpha$. Then $a_{i+1} \ldots a_n$ is an interference chain from $dom(a_i)$ to $u$ by definition. Then there is also an interference chain $\eta$ from $dom(a_i)$ to $u$ in $ipurge(\alpha, u)$ by Lem. 2, so $a_1 a_2 \ldots a_n \cdot \eta$ is an interference chain from $dom(a)$ to $u$ in $\gamma \cdot ipurge(\alpha, u)$. Therefore $ipurge(a \cdot \gamma \cdot ipurge(\alpha, u), u) = a \cdot ipurge(\gamma \cdot ipurge(\alpha, u), u)$. Then we have $ipurge(a \cdot \gamma \cdot \alpha, u) = ipurge(a \cdot \gamma \cdot ipurge(\alpha, u), u)$ by pre-pending action $a$ on both sides of the induction hypothesis.
- If $ipurge(a \cdot \gamma \cdot \alpha, u) = ipurge(\gamma \cdot \alpha, u)$, then there does not exist an interference chain in $\gamma \cdot \alpha$ from $dom(a)$ to $u$. Therefore there does not exist an interference chain in $\gamma \cdot ipurge(\alpha, u)$ which is a shorter sequence. Then we have $ipurge(a \cdot \gamma \cdot ipurge(\alpha, u), u) = ipurge(\gamma \cdot ipurge(\alpha, u), u)$. Then we have the result by induction hypothesis. ∎

Since the effect of $ipurge$ on the policy $\rightsquigarrow$ is the same as that of $purge_{\Pi(\rightsquigarrow)}$, every policy $\Pi(\rightsquigarrow)$ encoding an intransitive noninterference policy $\rightsquigarrow$ is right-consistent. Together with the unwinding characterization for policies of post-conditional assertions in Sect. IV, this result makes it possible to reason about security with respect to intransitive noninterference by unwinding theorems that are both

---

[8]Note that intransitive noninterference policies are not necessarily left-consistent, since a prefix of a sequence does not necessarily contain an interference chain even if the whole sequence does. However, intuitively, left-consistency is not important for intransitive policies which only place controls after an action is performed.

*sufficient* and *necessary*. This allows us to verify security properties that are related to intransitive information flow in a variety of areas (such as operating system and security protocol verification) in a more precise way.

Moreover, our policy language on post-conditional assertions is strictly more expressive than the policies of intransitive noninterference, even in the case of Part = $\{\mathcal{A}_u \mid u \in U\}$. An example could be a four-user system with $U = \{H, D_1, D_2, L\}$, on which we have a policy with a single assertion $\langle \mathcal{A}_H \not\rightsquigarrow L \; [\![ \Diamond \mathcal{A}_{D_1} \Diamond \mathcal{A}_{D_2} ]_{\rightarrow}^{post} ]\!] \rangle$, but neither $\mathcal{A}_{D_1}$ nor $\mathcal{A}_{D_2}$ is restricted from interfering with $L$. This policy asserts that an action from $H$ must be approved by both $D_1$ and $D_2$ in the particular order before being passed on to $L$, and $D_1$ is allowed to pass information to $L$ in a way independent to the actions from $D_2$. This policy is not expressible by intransitive noninterference. Moreover it is not hard to show that such policy is still right-consistent.

## IV. UNWINDING RELATIONS

Unwinding provides a verification technique on noninterference-related security requirements. An unwinding theorem reduces the verification of an information flow security problem into the existence of a set of relations satisfying certain properties, which is thus easier to be formalized and verified by existing tools such as proof assistants and model checkers.[9] In this section we present general forms of unwinding theorems for the two classes of conditional noninterference assertions introduced in the previous sections.

The use of unwinding relations on the proof of noninterference has been discussed in the literature [18], [32] which is based on the assumption that the relation $\rightsquigarrow \subseteq U \times U$ is transitive. First we show that this result is still valid for the class of policies that consist of strict assertions. (Note here the relation $\rightsquigarrow$ as determined by the set of assertions is not necessarily transitive.) Given a machine $M = \langle S, s_0, step, obs, O \rangle$ and a policy $\Pi$ consisting of only strict assertions, a set of unwinding relations $\{\sim_u\}_{u \in U}$ are defined as follows. For each user $u \in U$, $\sim_u \subseteq S \times S$ is an equivalence relation satisfying the conditions output consistency (OC), step consistency (SC), and local respect $\rightsquigarrow$ (LR).

OC $\quad s \sim_u t$ implies $obs_u(s) = obs_u(t)$.
SC $\quad s \sim_u t$ and $a \in A$ implies $step(s, a) \sim_u step(t, a)$.
LR $\quad s \sim_u step(s, a)$ if $\langle part(a) \not\rightsquigarrow u \rangle \in \Pi$.

The existence of a set of relations $\{\sim_u\}_{u \in U}$ that satisfy the above three properties is both sufficient and necessary for a system to be secure. The proof method is exactly the same as what was presented in [32]. Define a relation $\stackrel{u}{\sim} \subseteq S \times S$ for each $u \in U$ by $s \stackrel{u}{\sim} t$ if $obs_u(s) = obs_u(t)$.

*Theorem 1:* Given a policy $\Pi$, a system $M$ is secure with respect to $\Pi$ iff there exist unwinding relations $\{\sim_u\}_{u \in U}$.

*Proof:* The 'if' direction can be proved by induction on the length of the input actions in the same style of [32]. For the 'only if' direction, if the $M$ is secure, we can show that the relations $\stackrel{u}{\approx}$ defined by $s \stackrel{u}{\approx} t$ if $s \bullet \alpha \stackrel{u}{\sim} t \bullet \alpha$ for all $\alpha \in A^*$ satisfies OC, SC and LR. ∎

### A. Unwinding for Pre-conditional Assertions

We present an unwinding technique which is sound for policies consisting of pre-conditional assertions defined by the policy language $\Phi$. This technique is complete for policies that are left-consistent. Since the policy language produces a regular set of sequences, for each assertion $T$ of the form $\langle P \not\rightsquigarrow u \; [\![\phi^{pre}]\!]\rangle$, we write $A(\phi)^{P,u}$ for the finite automaton accepting $L(\phi)$, and regard $A(\phi)^{P,u}$ as the assertion automaton of $T$.

We define an additional rule for the unwinding relations on pre-conditional assertions. Given a machine $M$ in the form of $\langle S, s_0, step, obs, O \rangle$ and a policy $\Pi$, a set of unwinding relations $\{\sim_u\}_{u \in U}$ are equivalence relations satisfying OC, SC, LR, and the new condition $\text{LR}^{\leq}$ which is specified as follows.

$\text{LR}^{\leq}$ $\quad s \sim_u step(s, a)$ if $\langle part(a) \not\rightsquigarrow u \; [\![\phi^{pre}]\!]\rangle \in \Pi$ and there exists $\alpha \in L(\phi)$ such that $s = s_0 \bullet \alpha$.

As LR ensures a partition to follow a strict assertion, the condition $\text{LR}^{\leq}$ ensures the satisfaction of pre-conditional assertions in general. Intuitively, if a state is reachable by an action sequence within the language defined by an assertion, an action that is controlled by that assertion must be purged. We show that this characterization is sufficient for a system to be secure with respect to a policy consisting of only pre-conditional assertions. (As a strict assertion can also be treated as a pre-conditional assertion by the regular expression $\mathcal{A}^*$.)

*Theorem 2:* Given a system $M$ and a policy $\Pi$ consisting of only pre-conditional assertions, $M$ is secure if there exists a set of equivalence relations $\{\sim_u\}_{u \in U}$ satisfying OC, SC, LR and $\text{LR}^{\leq}$.

If a given policy is left-consistent, then this characterization is also complete.

*Theorem 3:* Given a system $M$ and a policy $\Pi$ consisting of only pre-conditional assertions, if $M$ is secure and $\Pi$ is left-consistent, then there exist a set of equivalence relations $\{\sim_u\}_{u \in U}$ satisfying OC, SC, LR and $\text{LR}^{\leq}$.

The regularity of the assertion language $\Phi$ allows to apply assertion automata for pre-conditional assertions to mark the states where $\text{LR}^{\leq}$ needs to be applied to purge an action. This could be done by a parallel composition of the machine $M$ with the $A(\phi)^{P,u}$ for every $\langle P \not\rightsquigarrow u \; [\![\phi^{pre}]\!]\rangle \in \Pi$, which could be automated in a model checker. Since assertion automata usually do not contain a lot of states, a local model checking algorithm is able to detect violations of security on-the-fly when a system is very large (even possibly of infinite states). We have the following reduction from noninterference security properties with policies consisting of pre-conditional assertions to safety properties.

---

[9] Although noninterference are trace-based properties and unwinding are bisimulation-based techniques, the unwinding characterizations in this paper are tight partially because for deterministic systems trace semantics and bisimulation semantics coincide [41]. Extending unwinding as a complete characterization for trace-based information flow properties in nondeterministic systems will be challenging, and we leave it as a future work.

For an assertion $T = \langle P \not\rightarrow u \; [\![\phi^{pre}]\!]\rangle \in \Pi$, we assume that an assertion automaton $A(\phi)^T = \langle S_T, s_{(T,0)}, \rightarrow, \mathcal{F}_T\rangle$ is deterministic, and accepts the language $L(\phi)$. We assume $\Pi$ is denumerable as $\{T_1, T_2, \ldots\}$. Given a machine $M = \langle S, s_0, step, obs, O\rangle$, for each $u \in U$, we define a machine $M_u^\Pi = \langle S^u, s_0^u, step^u, obs^u, dom\rangle$ to be the system with identical actions and domains, with states $S^u = S \times S \times S_{T_1} \times S_{T_2} \times \ldots$, initial state $s_0^u = (s_0, s_0, s_{(T_1,0)}, s_{(T_2,0)}, \ldots)$, and the observation function $obs^u : S^u \rightarrow (O \times O)$ is defined as $obs^u(s_1, s_2, t_1, t_2, \ldots) = (obs_u(s_1), obs_u(s_2))$ for $s_1, s_2 \in S$, and transition function $step^u : S^u \times \mathcal{A} \rightarrow S^u$ is given by $step^u((s_1, s_2, t_1, t_2, \ldots), a) = (s_1', step(s_2, a), t_1', t_2', \ldots)$ with $a \in \mathcal{A}$ and $t_i \xrightarrow{a} t_i'$ for all $i$, and
$$s_1' = \begin{cases} s_1 & \text{if there is } T_i = \langle part(a) \not\rightarrow u \; [\![\phi^{pre}]\!]\rangle \\ & \text{and } t_i \in \mathcal{F}_{T_i}, \\ step(s_1, a) & \text{otherwise.} \end{cases}$$

Intuitively, in every transition, an action $a$ is not allowed to apply on the left part of a state pair, if the assertion automaton controlling $part(a)$ and associated with $u$ is in its final state. A proof by induction shows that for every sequence of actions $\alpha \in A^*$, if $s_0^u \bullet \alpha = (s, t, \ldots)$ in $M_u^\Pi$, then in $M$ we have $s = s_0 \bullet purge_\Pi(\alpha, u)$ and $t = s_0 \bullet \alpha$. We therefore obtain the following.

**Proposition 4:** A machine $M$ is secure with respect to a left-consistent policy $\Pi$ iff for all $u \in U$ and for all states $s$ in $M_u^\Pi$ reachable from $s_0^u$, we have that $obs^u(s) = (o, o')$ implies $o = o'$.

*B. Unwinding for Post-conditional Assertions*

In this section we study the unwinding relations for policies consisting of post-conditional assertions defined by $\Phi^-$ as given in Fig. 2. The design of unwinding for this class of policies is rather involved. Our solution allows possibly more than one equivalence relations for each user. The underlying intuition is as follows. If an action $a$ is allowed to interfere with user $u$ only if it is followed by another action $b$, then for each state $s$, we need to have $s$ and $step(s, a)$ indistinguishable by $u$ after any sequence of actions that does not contain $b$. Based on that, we define a binary relation $\overset{[b]}{\sim}_u \subseteq S \times S$ and let $s \overset{[b]}{\sim}_u step(s, a)$ to represent the effect that state $s$ and state $step(s, a)$ are not distinguishable by $u$ as long as $b$ is not performed. (i.e., $s \overset{[b]}{\sim}_u t$ implies $step(s, c) \overset{[b]}{\sim}_u step(t, c)$ if $c \neq b$) Intuitively, such a relation must be an *equivalence* relation. For readability we move some of the proofs in this section into appendix and only provide explanations about the proofs instead.

Let $\Pi$ be a policy of post-conditional assertions. For a user $u \in U$, write the set of assertions associated with $u$ as a subpolicy $\Pi_u \subseteq \Pi$. Let $Q = \mathcal{P}(\texttt{Part}) \cup \{\Diamond \mathcal{C} \mid \mathcal{C} \subseteq \texttt{Part}\}$. Define the set of terms which are suffixes of the given constraints in $\Pi_u$ as $\Delta_u^\Pi = \{\lambda \in Q^* \mid \exists \lambda' \in Q^*, \langle P \not\rightarrow u \; [\![C_1 \cup C_2 \cup \cdots \cup C_n]_\rightarrow^{post}]\!]\rangle \in \Pi_u : \lambda' \cdot \lambda = C_i \wedge i \in [1 \ldots n]\}$. Intuitively, this is the suffix closure of the set of post-conditional channels that allow to downgrade information from some partition to $u$. The set of unwinding relations for a user $u \in U$ is $\{\overset{\delta}{\sim}_u \mid \delta \subseteq \Delta_u^\Pi\}$, which are the equivalence relations satisfying the following rules.

OC $s \overset{\delta}{\sim}_u t$ implies $s \overset{u}{\sim} t$ for all $\delta \subseteq \Delta_u^\Pi$ with $\delta \cap \{\epsilon\} = \emptyset$.

SC$^+$ If $s \overset{\delta}{\sim}_u t$ and $a \in \mathcal{A}$, then $step(s, a) \overset{sc(\delta, a)}{\sim}_u step(s, a)$.

LR $\langle part(a) \not\rightarrow u \rangle \in \Pi$ implies $s \overset{\emptyset}{\sim}_u step(s, a)$.

LR$^\geq$ $\langle part(a) \not\rightarrow u \; [\![\lambda_1 \cup \lambda_2 \cup \ldots \lambda_n]_\rightarrow^{post}]\!]\rangle \in \Pi$ implies $s \overset{\{\lambda_1, \lambda_2, \ldots \lambda_n\}}{\sim}_u step(s, a)$.

SUB For all $\delta_1, \delta_2 \in \mathcal{P}(\Delta_u^\Pi)$, $\delta_1 \subseteq \delta_2$ implies $\overset{\delta_1}{\sim} \subseteq \overset{\delta_2}{\sim}$.

The function $sc : \mathcal{P}(\Delta_u^\Pi) \times \mathcal{A} \rightarrow \mathcal{P}(\Delta_u^\Pi)$ is defined as $sc(\delta) = \bigcup_{\lambda \in \delta} cut(\lambda, a)$, where the *cut* function is defined as follows. Given $P \in \mathcal{P}(\texttt{Part})$ and $\lambda \in Q^*$,

- $cut(\epsilon, a) = \{\epsilon\}$ for all $a \in \mathcal{A}$,
- $cut(P \cdot \lambda, a) = \{\lambda\}$ if $a \in P$,
- $cut(P \cdot \lambda, a) = \emptyset$ if $a \notin P$,
- $cut(\Diamond P \cdot \lambda, a) = \{\lambda\}$ if $a \in P$,
- $cut(\Diamond P \cdot \lambda, a) = \{\Diamond P \cdot \lambda\}$ if $a \notin P$.

The condition OC asserts that all such relations containing unfinished downgrading channels to $u$ (with $\epsilon \in \delta$) must be contained in $\overset{u}{\sim}$, i.e., they shall not currently be distinguished by $u$. The definition of the SC$^+$ rule follows the mechanism of pattern matching which simulates the process of purging. For example, if $s \overset{\{\Diamond P \lambda\}}{\sim}_u t$, then after an action $a \in P$ is performed, $step(s, a)$ and $step(t, a)$ needs to be related by the relation $\overset{\{\lambda\}}{\sim}_u$, indicating that an action in $P$ has been performed and that the rest of the downgrading channel is $\lambda$. The states can be related by two downgrading channels, e.g. $s \overset{\{\lambda, \lambda'\}}{\sim}_u t$, indicating the two possibilities to effect the view (or to relax the indistinguishability relation) of $u$. When two states are related by a set with a completed channel, e.g., $s \overset{\delta}{\sim}_u t$ with $\epsilon \in \delta$, then $s$ and $t$ need not be indistinguishable to $u$ any more. Plainly $\overset{\delta}{\sim}_u = S \times S$ if $\epsilon \in \delta$, where $S$ is the state space of a machine. Informally, condition SUB indicates that the more channels a relation carries, the weaker policies that relation represents. As $\overset{\emptyset}{\sim}_u$ is the smallest such relation for user $u \in U$, it represents strict noninterference, so that $u$ can never distinguish two states that are related by his own future behaviours. For a suffix constraint $\lambda$ in the form of $C\lambda'$ or $\Diamond C\lambda'$, write $I(\lambda)$ for $C$ which is the first set of actions to check in $\lambda$. We have the following property for function $sc$.

**Lemma 4:** For all $\lambda \in sc(\delta, a)$, we have at least one of the following conditions hold.

1) $\lambda = \epsilon$ and $\lambda \in \delta$,
2) $\lambda \in \delta$ with $a \notin I(\lambda)$,

3) $C\lambda \in \delta$ with $a \in C$,
4) $\Diamond C\lambda \in \delta$ with $a \in C$.

**Lemma 5:** For all $\delta_1, \delta_2 \in \mathcal{P}(\Delta_u^\Pi)$ and $u \in U$, $s \stackrel{\delta_1}{\sim}_u t$ and $t \stackrel{\delta_2}{\sim}_u r$ implies $s \stackrel{\delta_1 \cup \delta_2}{\sim}_u r$.

*Proof:* By the rule SUB we have $\stackrel{\delta_1}{\sim} \subseteq \stackrel{\delta_1 \cup \delta_2}{\sim}$ and $\stackrel{\delta_2}{\sim} \subseteq \stackrel{\delta_1 \cup \delta_2}{\sim}$, therefore $s \stackrel{\delta_1 \cup \delta_2}{\sim}_u t$ and $t \stackrel{\delta_1 \cup \delta_2}{\sim}_u r$. Then $s \stackrel{\delta_1 \cup \delta_2}{\sim}_u r$ by transitivity of the relation $\stackrel{\delta_1 \cup \delta_2}{\sim}_u$. ∎

Similar to the pre-conditional constraints, every post-conditional constraint can be regarded as a pattern in regular expression, such that an action must not be purged if it is followed by a sequence of actions within the pattern characterized by the constraint. Define an interpretation operator $[.] : \Phi^- \to (\mathcal{P}(\mathcal{A}))^*$, by $[\epsilon] = \mathcal{A}^*$, $[C\lambda] = C[\lambda]$, and $[\Diamond C\lambda] = (\mathcal{A}\backslash C)^* C[\lambda]$ for $C \subseteq \mathcal{A}$, where $\lambda \in (\mathcal{P}(\mathcal{A}))^*$.

**Lemma 6:** Given a system $M$, a user $u \in U$, and a policy $\Pi$ with only post-conditional assertions, if there exists a set of relations $\{\stackrel{\delta}{\sim}_u\}_{\delta \subseteq \Delta_u^\Pi, u \in U}$ satisfying OC, LR, LR$^\geq$, SC$^+$ and SUB, then for all $s, t \in S$ and $\alpha \in A^* \setminus \bigcup_{\lambda \in \delta}[\lambda]$ with $s \stackrel{\delta}{\sim} t$ and $\delta \subseteq \Delta_u^\Pi$ satisfying $\delta \cap \{\epsilon\} = \emptyset$, we have $s \bullet \alpha \stackrel{u}{\sim} t \bullet purge_\Pi(\alpha, u)$.

The proof of this lemma is by induction on the length of an action sequence on states that are related by all possible sets of incomplete channels. From Lem. 6 one can obtain the soundness result.

**Theorem 4:** Given a system $M$, a user $u \in U$, and a policy $\Pi$ with only post-conditional assertions, if there exists a set of relations $\{\stackrel{\delta}{\sim}_u\}_{\delta \subseteq \Delta_u^\Pi, u \in U}$ satisfying OC, LR, LR$^\geq$, SC and SUB, then $M$ is secure with respect to $\Pi$.

*Proof:* We need to show for all $u \in U$ and $\alpha \in A^*$, we have $s_0 \bullet \alpha \stackrel{u}{\sim} s_0 \bullet purge_\Pi(\alpha, u)$. Since $\stackrel{\emptyset}{\sim}_u$ is reflexive we have $s_0 \stackrel{\emptyset}{\sim}_u s_0$, then the result directly follows by Lem. 6. Note $\bigcup_{\lambda \in \emptyset}[\lambda] = \emptyset$. ∎

To establish a completeness result, we study the set of relations $\{\stackrel{\delta}{\approx}_u\}_{\delta \in \Delta_u^\Pi}$ specified as follows. Define $\stackrel{\delta}{\approx}_u \subseteq S \times S$, such that $s \stackrel{\delta}{\approx}_u t$ if for all $\alpha \in \mathcal{A}^*$ satisfying $\alpha \notin [\lambda]$ for all $\lambda \in \delta$, $s \bullet \alpha \stackrel{u}{\sim} t \bullet \alpha$. We regard $\{\stackrel{\delta}{\approx}_u\}_{\delta \in \Delta_u^\Pi}$ as the relations that characterize information flow security for post-conditional assertions, with some nice properties that are guaranteed by Lem. 7.

**Lemma 7:** For each user $u \in U$ in system $M$, the set of relations $\{\stackrel{\delta}{\approx}_u\}_{\delta \in \Delta_u^\Pi}$ satisfies OC, SC$^+$ and SUB.

Finally we are able to prove that the existence of such unwinding relations is also necessary for a system to be secure, provided that the given policy consisting of post-conditional assertions is right-consistent. The methodology on proving Thm. 5 is that OC, SC$^+$ and SUB conditions determine a set of the largest bisimulation-like relations $\{\stackrel{\delta}{\approx}_u\}_{\delta \subseteq \Delta_u^\Pi}$ on the state space for each $u$, then LR and LR$^\geq$ conditions assert that noninterfering actions do not make transitions that go beyond each equivalent class. We leave the detailed proof in the appendix.

**Theorem 5:** Given a system $M$ with a right-consistent policy $\Pi$ consisting of post-conditional assertions, if $M$ is secure with respect to $\Pi$, then there exists a set of relations $\{\stackrel{\delta}{\sim}_u\}_{\delta \subseteq \Delta_u^\Pi}$ satisfying OC, LR, LR$^\geq$, SC$^+$ and SUB for all $u \in U$.

### C. A Case Study on Unwinding

We take the policy as introduced in example 2 and show how to construct unwinding relations in this simple system to ensure integrity of data-base operations. Suppose there are a finite number of employees $\mathbb{E} = \{E_1, E_2 \ldots E_m\}$ working with a database $\mathbb{B}$ with finite entries $X = \{x_1, x_2, \ldots x_n\}$ each of which stores a natural number. The action set available to $E_i$ is $A_{E_i}^r \cup A_{E_i}^w \cup \{a_{E_i}^{bk}\}$, where $\mathcal{A}_{E_i}^r = \{r(i, x) \mid x \in X\}$ and $\mathcal{A}_{E_i}^w = \{w(i, x, v) \mid x \in X\}$. The state space is $S = (\{succ, deny, ready, \bot\} \cup \mathbb{N})^\mathbb{E} \times \mathbb{N}^X$, so that a state $s = (o_1, o_2, \ldots o_m, d_1, d_2, \ldots d_n)$ is a snapshot of all employee's observations as well as the contents in database $\mathbb{B}$. In this case we write $s(i)$ for $E_i$'s observation and $s(x_j)$ for the $j$-th entry of $\mathbb{B}$ in $s$. The observation function for $\mathbb{B}$ (as a user) is thus $obs_\mathbb{B}(s) = (s(x_1), s(x_2), \ldots s(x_n))$, and $obs_{E_i}(s) = s(i)$. Write $s[t \mapsto v]$ for a state identical to $s$ except that $s[t \mapsto v](t) = v$. The initial state $s_0$ is defined as $s_0(i) = \bot$ for all $i \in 1 \ldots m$, and $s_0(x_j) = 0$ for all $x_j \in X$. The transition function is defined as follows. For all $i \in 1 \ldots m$ and $x_k \in X$,

- $step(s, r(i, x_k)) = s[s(i) \mapsto s(x_k)][\forall j \neq i : s(j) \mapsto \bot]$,
- $step(s, w(i, x_k, v)) = s[s(i) \mapsto deny][\forall j \neq i : s(j) \mapsto \bot]$ if $s[i] \neq ready$, and $step(s, w(i, x_k, v)) = s[s(i) \mapsto succ][s(x_k) = v][\forall i : s(i) \mapsto \bot]$ otherwise,
- $step(s, a_{E_i}^{bk}) = s[s(i) \mapsto ready][\forall j \neq i : s(j) \mapsto \bot]$.

where $[\forall j \neq i : s(i) \mapsto \bot]$ is short for $[s(1) \mapsto \bot] \ldots [s(i-1) \mapsto \bot][s(i+1) \mapsto \bot] \ldots [s(m) \mapsto \bot]$, which sets all users except $i$'s observation to $\bot$. Informally, $a_{E_i}^{bk}$ acquires a unique *write-permission* for $E_i$ by setting $i$'s observation to $ready$ and simultaneously removing all other employees' ability to write.

Recalling the security policy of example 2, we have the following three rules to ensure integrity of $\mathbb{B}$. For all $E_i$, (1) reading actions do not modify $\mathbb{B}$: $\langle \mathcal{A}_{E_i}^r \not\hookrightarrow \mathbb{B} \rangle$, (2) writing actions take effect only by immediately following a book-keeping action: $\langle \mathcal{A}_{E_i}^w \not\hookrightarrow \mathbb{B} \; [\![a_{E_i}^{bk}]_\searrow^{pre}]\!] \rangle$, and (3) book-keeping does not have side effects: $\langle \{a_{E_i}^{bk}\} \not\hookrightarrow \mathbb{B} \; [\![\mathcal{A}_{E_i}^w]_\to^{post}]\!] \rangle$.

We treat (1) and (2) as pre-conditional assertions, by defining an equivalence relation $\sim_\mathbb{B}$ as follows.[10] Let $s \sim_\mathbb{B} t$ if $obs_\mathbb{B}(s) = obs_\mathbb{B}(t)$ and for all $1 \leq i \leq m$, either $s(i) = t(i) = ready$, or $s(i) \neq ready$ and $t(i) \neq ready$.

---

[10]Since the policy is not designed to protect the employees, we only study the unwinding relations for $\mathbb{B}$.

We show that $\sim_\mathbb{B}$ is an unwinding relation for assertions (1) and (2).

- OC is trivial.
- For SC, if $s \sim_\mathbb{B} t$, then for all $1 \leq i \leq m$, (1) $step(s, r(i,x)) \sim_\mathbb{B} step(t, r(i,x))$, because $r(i,x)$ only sets $E_i$'s observation to $s(x)$ which is the same as $t(x)$ by definition, and (2) $step(s, w(i,x,v)) \sim_\mathbb{B} step(t, w(i,x,v))$, since the writing action either changes both item $x$ to $v$, or fails to change both, and (3) $step(s, a_{E_i}^{bk}) \sim_\mathbb{B} step(t, a_{E_i}^{bk})$, since the book-keeping action only sets both states as ready for $E_i$ to write, and resets all other observations to $\bot$.
- For LR, it is obvious that $s \sim_\mathbb{B} step(s, r(i,x))$ for all $i$ and $x$.
- For LR$^\leq$, the language $L([a_{E_i}^{bk}]_\searrow^{pre})$ is expressed as $\mathcal{A}^*(\mathcal{A} \setminus \{a_{E_i}^{bk}\})$. Then we have that for all $\alpha \in \mathcal{A}^*(\mathcal{A} \setminus \{a_{E_i}^{bk}\})$ and action $a$ in the form of $w(i,x,v)$, $step(s_0 \bullet \alpha, a) \sim_\mathbb{B} s_0 \bullet \alpha$ (Since no one is ready in $s_0 \bullet \alpha$ and no one is ready in $step(s_0 \bullet \alpha, a)$).

Assertion (3) is post-conditional, for which we establish the following relations.

- $\overset{\{\epsilon\}}{\sim}_\mathbb{B} = \overset{\{A_{E_i}^w, \epsilon\}}{\sim} = S \times S$ for all $i$.
- $s \overset{\{A_{E_i}^w\}}{\sim}_\mathbb{B} t$ if $obs_\mathbb{B}(s) = obs_\mathbb{B}(t)$, and for all $j \neq i$, either $s(j) = t(j) = ready$, or $s(j) \neq ready$ and $t(j) \neq ready$. (i.e., only $E_i$'s observation is relaxed from the constraints imposed on $\sim_\mathbb{B}$.)
- $\overset{\emptyset}{\sim}_\mathbb{B}$ is defined as $\sim_\mathbb{B}$.

We show this set of relations are unwinding relations for assertion (3).

- OC and SUB are trivial.
- For SC$^+$, the case for $\overset{\delta}{\sim}_\mathbb{B}$ with $\epsilon \in \delta$ is trivial, since in this case $\overset{\delta}{\sim}_\mathbb{B} = S \times S$. Let $s \overset{\{A_{E_i}^w\}}{\sim}_\mathbb{B} t$, then for all $a \in \mathcal{A} \setminus \mathcal{A}_{E_i}^w$, we need to show $step(s,a) \overset{\emptyset}{\sim}_\mathbb{B} step(t,a)$. This is straightforward because the only possibility to prevent $s$ to be related to $t$ by $\overset{\emptyset}{\sim}_\mathbb{B}$ is that they disagree on $E_i$'s observation, and every action $a \in \mathcal{A} \setminus \mathcal{A}_{E_i}^w$ will set $E_i$'s observation to the same value in $step(s,a)$ and $step(t,a)$ without modifying $\mathbb{B}$'s contents.
- For LR$^\geq$, for all $s \in S$, only $E_i$'s view is changed to $ready$ in $step(s, a_{E_i}^{bk})$, thus $s \overset{\{A_{E_i}^w\}}{\sim}_\mathbb{B} step(s, a_{E_i}^{bk})$ by definition.

By establishing the unwinding relations, Thm. 2 and Thm. 4 guarantee that the system is secure with respect to the given policy. Moreover, one can still prove that the existence of such unwinding relations is complete for this particular policy in this example, by applying the techniques used in the proofs of Thm. 3 and Thm. 5. As the policy discussed in this example is neither left-consistent nor right-consistent (which can be shown from the $purge$ function derived from the policy)[11], this serves as an example showing that left- and right-consistencies are not always necessary for a policy to be completely characterizable by the unwinding relations defined in this paper.

## V. RELATED WORK

Conditional noninterference was first proposed by Goguen and Meseguer [17]. Our work extends their definition to a more general form, such that the control of information flow can be placed either before or after the actions with intended flow. The notion of intransitive noninterference was first proposed by Haigh and Young [21], and later revised by Rushby [32]. Our policy defined by post-conditional assertions are strictly more expressive than that of intransitive noninterference, which has been sketched in Sect. III-E. Nevertheless, the unwinding theorems presented for this more general policy is both sound and complete in a very general sense (we believe that action-based channel control policies are usually supposed to be consistent), while the weak unwinding relation [32] fails to be complete for intransitive noninterference in the literature. The unwinding technique of Mantel [23] is sound for a spectrum of trace-based properties [22], but it is also not complete. A few other works extend Rushby's weak unwinding in nondeterministic language-based settings [25], [24]. The result in this paper is based on systems with deterministic transition functions, but it will be straightforward to extend the definitions of our policies for both pre-conditional assertions and post-conditional assertions in nondeterministic systems, possibly by revising the unwinding rule SC (or SC$^+$) in the way of bisimulation [27][12].

Bossi et al. extended the unwinding-based characterization for the security properties in SPA [14], [16] to support downgrading [6]. They described a policy for three security levels including $H$ (High level user), $D$ (Downgrader) and $L$ (low level user) by applying unwinding to disallow information flow from $H$ to $L$ without putting any constraint on $D$. Their approach is basically Goguen and Meseguer's strict noninterference policies [17] (as we sketched in example 1) in a nondeterministic environment with silent system moves. With respect to persistency [16], our policies by post-conditional assertions are inherently persistent, i.e., if a system is secure with respect to such policies then it is secure if every reachable state is a possible initial state. However, our policies by pre-conditional assertions are not necessarily persistent by definition.[13] Roscoe and Goldsmith [31] generalized the determinism based notion of noninterference [30] to intransitive noninterference with three security levels in process algebra CSP. Their property is potentially stronger than most of the existing intransitive noninterference properties in literature [36].

---
[11]Nevertheless, it is obvious that the pre-conditional part of the policy is left-consistent, and the post-conditional part of the policy is right-consistent, which helps to establish a proof for completeness.

[12]However, achieving completeness might be very nontrivial for unwinding in nondeterministic systems for trace-based properties.

[13]In this case, we claim that it is sufficient to verify the persistent version of a pre-conditional assertion $\phi$ by examining a policy automaton accepting the language $L(\phi)' = \{\alpha \cdot \alpha' \mid \alpha \in \mathcal{A}^* \land \alpha' \in L(\phi)\}$.

Van der Meyden developed a new set of intransitive noninterference properties to reason about information flow epistemically [38]. As it was identified that Haigh and Young's intransitive flow property [21] may allow a downgrader to pass information from high level to low level without knowing what is to be downgraded, a number of new intransitive noninterference properties are introduced to catch the idea that a downgrader's knowledge about the secret information should be no less than what the low level user is able to get. The new properties defined by van der Meyden are stronger than intransitive noninterference [21], [32] and weaker than (strict) noninterference [17]. Our framework lies in a different dimension, in that we extend the framework of [17], [32] to support more flexible policies without much concern on a user's knowledge.

The methodologies for declassification of secret information have been surveyed by Sabelfeld and Sands [34], in which all related works are classified into four different dimensions: (1) *who* releases information, (2) *what* information is released (3) *where* in the system information is released and (4) *when* information can be released. Although most of the surveyed works are in the language-based setting, the classification seems to make sense in the state-based models as well. Our policy design supports the *who* dimension, by assigning a partition of a particular user in a policy to control information flow to a user, and also the *where* and *when* dimensions, by controlling information release only after a downgrading channel is fully established (such as allowed by post-conditional assertions). In terms of flexibility, as this framework does not assume a centralized security policy, it is possible to express *integrity* for decentralized flow control [28], by assigning users privileged actions to switch on and off writing permits to the files they own. However as our policy is action-based, it might not be convenient to express decentralized *confidentiality* policies. More recently, Chong and Myers [8] define declassification and erasure policies that specify conditions under which information may be downgraded, or must be erased. Instead of binding policies on information, our pre-conditional policies focus on the control over the *source* and *destination* of information flow, by adding and removing permits from an action partition of a user via controlling actions.

Hadj-Alouane et al. studied verification of intransitive noninterference property in finite state systems [20]. In order to verify the property, they reduce a system into an automaton accepting the reversed language, which potentially consumes space exponential to the size of the system. Pinsky also proposed an algorithm to verify noninterference properties [29]. However that algorithm only works for transitive policies, but fails when the underlying information flow relation is intransitive. A new algorithm for intransitive noninterference is proposed by van der Meyden [37] which has a complexity bound polynomial to the size of a machine but exponential to the number of users. Verification on our unwinding relations for post-conditional assertions can be done in-place, therefore it is also polynomial time to the size of a system, but it could be exponential to the size of a policy (as shown in the subset construction on the set of post-conditional assertions when constructing unwinding relations). Nevertheless, our policies are strictly more general than intransitive noninterference policies, as shown in Sect. III-E. It will also be interesting to investigate algorithmic verification methods on generating unwinding relations in more general systems (i.e., systems that are not necessarily finite state), as it has been shown that verification of Mantel's BSPs [22] in push-down systems is undecidable [13]. The methodologies on reducing information flow properties to safety by self-composition have been discussed in the literature [2], [35], [11], [40], for a variety of system models.

## VI. Conclusion and Future Work

This work introduces a framework of information flow policies by noninterference assertions which generalizes existing work of both transitive and intransitive noninterference. Although noninterference is in general defined as a static security notion, we applied our policy language to express a number of dynamic security requirements including upgrading, downgrading and channel control. Our unwinding theorems on both pre-conditional and post-conditional assertions are novel, and they are more precise and more general than the existing results in the literature, to our knowledge.

There is a possible future direction to extend our policy by allowing clock *tick*s to act as upgrading or downgrading channels. This will make it possible to express *time-based control* in real-time systems, which might be an interesting future work to explore both upgrading and downgrading in the *when* dimension of [34].

There are plenty of extensions of noninterference in non-deterministic and probabilistic systems, and this will be an interesting future work for conditional noninterference. Also we believe that it will be of interest to find real cases where our unwinding theorems (or any suitable extensions) can be applied to verify their corresponding security requirements in more general systems. Furthermore, it is also possible to enrich our policy by incorporating state information into the policy language in a concrete system verification. Again, this will be of more interest in a sensible case study in the future.

## VII. Acknowledgement

The author thanks Peter Ryan for his useful comments on an earlier draft of the paper.

APPENDIX

This appendix contains the proofs of some results presented in the article.

*Proof:* (of Prop. 2) We prove by induction on the length of an action sequence. Base case: $ipurge(\epsilon, u) = purge_{\Pi(\leadsto)}(\epsilon, u) = \epsilon$.

Suppose $ipurge(\alpha, u) = purge_{\Pi(\leadsto)}(\alpha, u)$ for some $\alpha \in \mathcal{A}^*$, we show the case for $a \cdot \alpha$ with $a \in \mathcal{A}$. If $\alpha$ contains an interference chain from $dom(a)$ to $u$, then $ipurge(a \cdot \alpha, u) = a \cdot ipurge(\alpha, u)$. Also $purge_{\Pi(\leadsto)}(a \cdot \alpha, u) = a \cdot purge_{\Pi(\leadsto)}(\alpha, u)$, since the purging of the actions in $\alpha$ does not depend on $a$, and $\Pi(\leadsto)$ contains a condensed interference chain from $dom(a)$ to $u$ by definition. If $\alpha$ does not contain an interference chain from $dom(a)$ to $u$, then $ipurge(a \cdot \alpha, u) = ipurge(\alpha, u)$ and $purge_{\Pi(\leadsto)}(a \cdot \alpha, u) = purge_{\Pi(\leadsto)}(\alpha, u)$. In both cases we have $ipurge(a \cdot \alpha, u) = purge_{\Pi(\leadsto)}(a \cdot \alpha, u)$. ∎

*Proof:* (of Lem. 2) The 'if' direction is trivial, since $ipurge(\alpha, v)$ is contained in $\alpha$. For the 'only if' direction, suppose $\alpha' = a_1 a_2 \ldots a_n$ is an interference chain from $u$ to $v$ in $\alpha$, it can be shown by induction on every suffix of $\alpha'$ that all actions in $\alpha'$ will stay in the sequence $ipurge(\alpha, v)$. ∎

*Proof:* (of Lem. 3) By induction on length of $\alpha$. Base case is trivial.

Suppose $ipurge(ipurge(\alpha, u), u) = ipurge(\alpha, u)$, we show the case for $a \cdot \alpha$.

- If $\alpha$ contains an interference chain from $dom(a)$ to $u$, then $ipurge(\alpha, u)$ also contains an interference chain from $dom(a)$ to $u$ by Lem. 2. Therefore we have $ipurge(ipurge(a \cdot \alpha, u), u) = ipurge(a \cdot ipurge(\alpha, u), u) = a \cdot ipurge(ipurge(\alpha, u), u)$ and $ipurge(a \cdot \alpha, u) = a \cdot ipurge(\alpha, u)$. Since $ipurge(\alpha, u) = ipurge(ipurge(\alpha, u), u)$, we get $ipurge(a \cdot \alpha, u) = ipurge(ipurge(a \cdot \alpha, u), u)$ by induction hypothesis.
- If $\alpha$ does not contain an interference chain from $dom(a)$ to $u$, then $ipurge(\alpha, u)$ also does not contain an interference chain from $dom(a)$ to $u$ by Lem. 2. Therefore $ipurge(ipurge(a \cdot \alpha, u), u) = ipurge(ipurge(\alpha, u), u)$ and $ipurge(a \cdot \alpha, u) = ipurge(\alpha, u)$. Then we have $ipurge(a \cdot \alpha, u) = ipurge(ipurge(a \cdot \alpha, u), u)$ by induction hypothesis.

∎

*Proof:* (of Thm. 2) We show that if there exist relations $\{\sim_u\}_{u \in U}$ satisfying OC, SC, LR and LR$^{\leq}$, then for all $u \in U$, $\alpha \in A^*$, $s_0 \bullet \alpha \sim_u s_0 \bullet purge_\Pi(\alpha, u)$, then by OC we will have $s_0 \bullet \alpha \stackrel{u}{\sim} s_0 \bullet purge_\Pi(\alpha, u)$. we prove this by induction on the length of the action sequences. For $\alpha = \epsilon$, $purge_\Pi(\alpha) = \alpha = \epsilon$, we have $s_0 \sim_u s_0$ by the fact that $\sim_u$ is reflexive. Suppose for some $\alpha \in A^*$ we have $s_0 \bullet \alpha \sim_u s_0 \bullet purge_\Pi(\alpha, u)$, we show the case for $\alpha \cdot a$.

- If $purge_\Pi(\alpha \cdot a, u) = purge_\Pi(\alpha, u)$, we have the following two cases: (1) $\langle part(a) \not\leadsto u \rangle \in \Pi$, (2) $\langle part(a) \not\leadsto u \llbracket \phi^{pre} \rrbracket \rangle \in \Pi$ and $\alpha \in L(\phi)$. In both cases we have $s_0 \bullet \alpha \sim_u s_0 \bullet (\alpha \cdot a)$ by LR (or LR$^{\leq}$). With the induction hypothesis $s_0 \bullet \alpha \sim_u s_0 \bullet purge_\Pi(\alpha \cdot a, u)$, by transitivity of $\sim_u$, we have $s_0 \bullet (\alpha \cdot a) \sim_u s_0 \bullet purge_\Pi(\alpha \cdot a, u)$.
- Otherwise, we have $purge_\Pi(\alpha \cdot a, u) = purge_\Pi(\alpha, u) \cdot a$. Then by the induction hypothesis and SC, we have $s_0 \bullet (\alpha \cdot a) \sim_u s_0 \bullet (purge_\Pi(\alpha, u) \cdot a)$, therefore $s_0 \bullet (\alpha \cdot a) \sim_u s_0 \bullet purge_\Pi(\alpha \cdot a, u)$.

∎

*Proof:* (of Thm. 3) Suppose $M$ is secure, we show that the relations $\{\sim_u\}_{u \in U}$ defined by $s \sim_u t$ if for all $\alpha \in A^*$, $s \bullet \alpha \overset{u}{\sim} t \bullet \alpha$ satisfy OC, SC, LR and LR$^\leq$.

- For OC, let $\alpha = \epsilon$, then we have $s \sim_u t$ implies $s \overset{u}{\sim} t$.
- For SC, let $s \sim_u t$ and $a \in \mathcal{A}$, if $step(s,a) \not\sim_u step(t,a)$, then there exists $\alpha \in \mathcal{A}^*$ such that $step(s,a) \bullet \alpha \overset{u}{\not\sim} step(t,a) \bullet \alpha$, then $s \bullet (a \cdot \alpha) \overset{u}{\not\sim} t \bullet (a \cdot \alpha)$ which contradicts $s \sim_u t$. Therefore $step(s,a) \sim_u step(t,a)$.
- For LR$^\leq$, let $\langle P \not\rightarrow u \, [\![\phi^{pre}]\!]\rangle \in \Pi$. If there exists $a \in P$ and $s \in S$ such that $s_0 \bullet \alpha$, $\alpha \in L(\phi)$ and $s \not\sim_u step(s,a)$, then there exists $\alpha' \in \mathcal{A}^*$ such that $s_0 \bullet \alpha \bullet \alpha' \overset{u}{\not\sim} step(s_0 \bullet \alpha, a) \bullet \alpha'$, which is equivalent to that $s_0 \bullet (\alpha \cdot \alpha') \overset{u}{\not\sim} s_0 \bullet (\alpha \cdot a \cdot \alpha')$. However, since policy $\Pi$ is left-consistent, we have $purge_\Pi(\alpha \cdot \alpha', u) = purge_\Pi(purge_\Pi(\alpha, u) \cdot \alpha', u)$, and $purge_\Pi(\alpha \cdot a \cdot \alpha', u) = purge_\Pi(purge_\Pi(\alpha \cdot a, u) \cdot \alpha', u)$. Then $purge_\Pi(\alpha \cdot \alpha', u) = purge_\Pi(\alpha \cdot a \cdot \alpha', u)$ by $purge_\Pi(\alpha, u) = purge_\Pi(\alpha \cdot a, u)$, i.e., $\alpha \cdot \alpha'$ and $\alpha \cdot a \cdot \alpha'$ have the same purged result with respect to $u$. Therefore we have either $s_0 \bullet (\alpha \cdot \alpha') \overset{u}{\not\sim} s_0 \bullet purge_\Pi(\alpha \cdot \alpha', u)$, or $s_0 \bullet (\alpha \cdot a \cdot \alpha') \overset{u}{\not\sim} s_0 \bullet purge_\Pi(\alpha \cdot a \cdot \alpha', u)$, contradicting the assumption that $M$ is secure.
- The case of LR is similar to LR$^\leq$.

∎

*Proof:* (of Lem. 6) We prove by induction on length of $\alpha$. Base case: $\alpha = \epsilon$, then $purge_\Pi(\epsilon, u) = \epsilon$, we have $s \overset{\delta}{\sim}_u t$ implies $s \overset{u}{\sim} t$ by OC for every $\delta \cap \{\epsilon\} = \emptyset$. Suppose this holds for an action sequence $\alpha$ on all states $s, t$, $\delta \subseteq \Delta_u^\Pi$ with $\delta \cap \{\epsilon\} = \emptyset$, such that $s \overset{\delta}{\sim}_u t$ with $\alpha \in \mathcal{A}^* \setminus \bigcup_{\lambda \in \delta}[\lambda]$, we show the case for $a \cdot \alpha$. Let $s \overset{\delta}{\sim}_u t$ with $a \cdot \alpha \in \mathcal{A}^* \setminus \bigcup_{\lambda \in \delta}[\lambda]$ and $\delta \cap \{\epsilon\} = \emptyset$.

- If $purge_\Pi(a \cdot \alpha, u) = a \cdot purge_\Pi(\alpha, u)$, then we have $step(s,a) \overset{sc(\delta,a)}{\sim}_u step(t,a)$. First we show that $\epsilon \notin sc(\delta, a)$. Because if $\epsilon \in sc(\delta, a)$, then by Lem. 4, either (1) $\epsilon \in \delta$, or (2) there is $C$ or $\Diamond C$ in $\delta$ such that $a \in [C]$ or $a \in [\Diamond C]$, which implies $a \cdot \alpha \in [C]$ or $a \cdot \alpha \in [\Diamond C]$. Case (1) contradicts the assumption that $\delta \cap \{\epsilon\} = \emptyset$, and case (2) contradicts the assumption that $a \cdot \alpha \in \mathcal{A}^* \setminus \bigcup_{\lambda \in \delta}[\lambda]$. Next we show for all $\lambda \in sc(\delta, a)$, $\alpha \notin [\lambda]$. Because if there were $\lambda \in sc(\delta, a)$ such that $\alpha \in [\lambda]$, by Lem. 4, we would have the following cases: (1) if $\lambda \in \delta$ with $a \notin I(\lambda)$, then $a \cdot \alpha \in [\lambda]$; (2) if $C\lambda \in \delta$ or $\Diamond C\lambda \in \delta$ with $a \in C$, then $a \cdot \alpha \in [C\lambda]$ or $a \cdot \alpha \in [\Diamond C\lambda]$. Both cases contradict the assumption that $a \cdot \alpha \in \mathcal{A}^* \setminus \bigcup_{\lambda \in \delta}[\lambda]$. Therefore for all $\lambda \in sc(\delta, a)$, $\alpha \notin [\lambda]$, i.e., $\alpha \in \mathcal{A}^* \setminus \bigcup_{\lambda \in sc(\delta,a)}[\lambda]$. Then by induction hypothesis, we have $step(s,a) \bullet \alpha \overset{u}{\sim} step(t,a) \bullet purge_\Pi(\alpha, u)$. Therefore $s \bullet (a \cdot \alpha) \overset{u}{\sim} t \bullet purge_\Pi(a \cdot \alpha, u)$.
- If $purge_\Pi(a \cdot \alpha, u) = purge_\Pi(\alpha, u)$, we have the following two cases.
  - If $\langle part(a) \not\rightarrow u \rangle \in \Pi$, then by LR, we have $s \overset{\emptyset}{\sim} step(s,a)$, then $step(s,a) \overset{\delta}{\sim} t$ by Lem. 5. By induction hypothesis, we have $step(s,a) \bullet \alpha \overset{u}{\sim} t \bullet purge_\Pi(\alpha, u)$, then we have $s \bullet (a \cdot \alpha) \overset{u}{\sim} t \bullet purge_\Pi(a \cdot \alpha, u)$.
  - If $\langle part(a) \not\rightarrow u \, [\![\lambda_1 \cup \lambda_2 \cup \cdots \cup \lambda_n]_\rightarrow^{post}]\!]\rangle \in \Pi$ and $\alpha \notin [\lambda_i]$ for all $i \in [1 \ldots n]$. By LR$^\geq$, we have $step(s,a) \overset{\{\lambda_1,\lambda_2,\ldots\lambda_n\}}{\sim}_u s$. Then we have $step(s,a) \overset{\delta \cup \{\lambda_1,\lambda_2,\ldots\lambda_n\}}{\sim} t$ by Lem. 5. Since $\alpha \notin \lambda'$ for all $\lambda' \in \delta$, and $\alpha \notin [\lambda_i]$ for all $i$ by assumption, we have $\alpha \in \mathcal{A}^* \setminus \bigcup_{\lambda' \in \delta \cup \{\lambda_1,\lambda_2,\ldots\lambda_n\}}[\lambda']$. Then by induction hypothesis, we get $step(s,a) \bullet \alpha \overset{u}{\sim} t \bullet (purge_\Pi(\alpha, u))$, therefore $s \bullet (a \cdot \alpha) \overset{u}{\sim} t \bullet purge_\Pi(a \cdot \alpha, u)$.

∎

*Proof:* (of Lem. 7)

- For OC, let $\delta \subseteq \Delta_u^\Pi$ be a set that does not contain $\epsilon$ and $s \overset{\delta}{\approx}_u t$. Since $\epsilon \notin [\lambda]$ for all $\lambda \in \Delta_u^\Pi \setminus \{\epsilon\}$, we have $s \bullet \epsilon \overset{u}{\sim} t \bullet \epsilon$. Therefore $s \overset{u}{\sim} t$.
- For SUB, let $s \overset{\delta}{\approx}_u t$ and $\delta \subseteq \delta' \subseteq \Delta_u^\Pi$, we need to show $s \overset{\delta'}{\approx} t$. Since $\delta \subseteq \delta'$, we have $\bigcup_{\lambda \in \delta}[\lambda] \subseteq \bigcup_{\lambda \in \delta'}[\lambda]$, so $\mathcal{A}^* \setminus \bigcup_{\lambda \in \delta'}[\lambda] \subseteq \mathcal{A}^* \setminus \bigcup_{\lambda \in \delta}[\lambda]$. By $s \overset{\delta}{\approx}_u t$, we have $s \bullet \alpha \overset{u}{\sim} t \bullet \alpha$ for all $\alpha \in \mathcal{A}^* \setminus \bigcup_{\lambda \in \delta}[\lambda]$, so $s \bullet \alpha \overset{u}{\sim} t \bullet \alpha$ for all $\alpha \in \mathcal{A}^* \setminus \bigcup_{\lambda \in \delta'}[\lambda]$. Then we have $s \overset{\delta'}{\approx} t$ by definition.
- For SC$^+$, let $s \overset{\delta}{\approx}_u t$ and $a \in \mathcal{A}$, we study the following cases on the relations which may relate $step(s,a)$ and $step(t,a)$.
  - If $\epsilon \in \delta$, then by definition $\overset{\{\epsilon\}}{\approx}_u = S \times S$, therefore $step(s,a) \overset{\{\epsilon\}}{\approx}_u step(t,a)$.
  - If $P\lambda \in \delta$ and $a \in P$, then for all $\alpha \in \mathcal{A}^* \setminus [\lambda]$, $step(s,a) \bullet \alpha \overset{u}{\sim} step(t,a) \bullet \alpha$, because if not, then we would have $s \bullet (a \cdot \alpha) \overset{u}{\not\sim} t \bullet (a \cdot \alpha)$ such that $a \cdot \alpha \notin [P\lambda]$. Therefore $step(s,a) \overset{\{\lambda\}}{\approx}_u step(t,a)$.
  - If $P\lambda \in \delta$ and $a \notin P$, then we have $step(s,a) \bullet \alpha \overset{u}{\sim} step(t,a) \bullet \alpha$ for all $\alpha \in \mathcal{A}^*$, because if not, then we would have $s \bullet (a \cdot \alpha) \overset{u}{\not\sim} t \bullet (a \cdot \alpha)$ such that $a \cdot \alpha \notin [P\lambda]$. Therefore $step(s,a) \overset{\emptyset}{\approx}_u step(t,a)$.
  - If $\Diamond P\lambda \in \delta$ and $a \in P$, then we have $step(s,a) \bullet \alpha \overset{u}{\sim} step(t,a) \bullet \alpha$ for all $\alpha \in \mathcal{A}^* \setminus [\lambda]$, which is similar to the case of $P\lambda \in \delta$. Therefore $step(s,a) \overset{\{\lambda\}}{\approx}_u step(t,a)$.
  - If $\Diamond P\lambda \in \delta$ and $a \notin P$, then we have $step(s,a) \bullet \alpha \overset{u}{\sim} step(t,a) \bullet \alpha$ for all $\alpha \in \mathcal{A}^* \setminus [\Diamond P\lambda]$, because if not, then we would have $s \bullet (a \cdot \alpha) \overset{u}{\not\sim} t \bullet (a \cdot \alpha)$ with $a \cdot \alpha \notin [\Diamond P\lambda]$. Therefore $step(s,a) \overset{\{\Diamond P\lambda\}}{\approx}_u step(t,a)$.

The above cases give us $cut(\lambda, a)$ for every member $\lambda \in \delta$. By SUB we take the union of all the single-ton and empty sets to get $(step(s,a), step(t,a)) \in \bigcup_{\lambda \in \delta} cut(\lambda, a)$. Therefore $step(s,a) \overset{sc(\delta,a)}{\sim} step(t,a)$ by definition.

∎

*Proof:* (of Thm. 5) Suppose $M$ is secure with respect to $\Pi$, then for each $u \in U$ the relation $\{\overset{\delta}{\approx}_u\}_{\delta \subseteq \Delta_u^\Pi}$ satisfy OC, SC$^+$ and SUB by Lem. 7. Then we only need to show they also satisfy LR and LR$^\geq$ in the following cases.

- Suppose the relations do not satisfy LR for some $u \in U$, then there exists a reachable state $s$ and an assertion $\langle part(a) \not\rightsquigarrow u \rangle \in \Pi$ such that $step(s,a) \overset{\emptyset}{\not\approx}_u s$. Therefore there exists some $\alpha \in \mathcal{A}^*$ such that $step(s,a) \overset{y}{\not\sim} s$. Since $s$ is reachable we have $s = s_0 \bullet \alpha'$ for some $\alpha' \in \mathcal{A}^*$. Then we have $s_0 \bullet (\alpha' \cdot a \cdot \alpha) \overset{y}{\not\sim} s_0 \bullet (\alpha' \cdot \alpha)$. However $purge_\Pi(\alpha' \cdot a \cdot \alpha, u) = purge_\Pi(\alpha' \cdot purge_\Pi(a \cdot \alpha, u), u)$, and $purge_\Pi(\alpha' \cdot \alpha, u) = purge_\Pi(\alpha' \cdot purge_\Pi(\alpha, u), u)$ by right-consistency of $\Pi$. Since $purge_\Pi(a \cdot \alpha, u) = purge_\Pi(\alpha, u)$ by $\langle part(a) \not\rightsquigarrow u \rangle \in \Pi$, we have $purge_\Pi(\alpha' \cdot a \cdot \alpha, u) = purge_\Pi(\alpha' \cdot \alpha, u)$. By the assumption that $M$ is secure, we have $s_0 \bullet (\alpha' \cdot a \cdot \alpha) \overset{u}{\sim} s_0 \bullet purge_\Pi(\alpha' \cdot a \cdot \alpha, u)$ and $s_0 \bullet (\alpha' \cdot \alpha, u) \overset{u}{\sim} s_0 \bullet (\alpha' \cdot \alpha)$. Then we have $s_0 \bullet (\alpha' \cdot a \cdot \alpha) \overset{u}{\sim} s_0 \bullet (\alpha' \cdot \alpha)$, which is contradiction. Therefore we have the relations $\{\overset{\delta}{\approx}_u\}_{\delta \subseteq \Delta_u^\Pi}$ satisfying LR for all $u \in U$.

- Suppose the relations do not satisfy LR$^\geq$, then there exists a reachable state $s$ and an assertion $\langle part(a) \not\rightsquigarrow u \; [\![\lambda]_\rightarrow^{post}]\!] \rangle$ such that $step(s,a) \overset{\{\lambda\}}{\not\approx}_u s$. So there exists $\alpha \in \mathcal{A}^* \setminus [\lambda]$, such that $s \bullet (a \cdot \alpha) \overset{y}{\not\sim} s \bullet \alpha$. Since $s$ is reachable, there exists $\alpha' \in A^*$ such that $s_0 \bullet \alpha' = s$. Therefore we have $s_0 \bullet (\alpha' \cdot a \cdot \alpha) \overset{y}{\not\sim} s_0 \bullet (\alpha' \cdot \alpha)$. Also since $\alpha \in \mathcal{A}^* \setminus [\lambda]$, by definition $purge_\Pi(a \cdot \alpha, u) = purge_\Pi(\alpha, u)$. Then we have $purge_\Pi(\alpha' \cdot a \cdot \alpha, u) = purge(\alpha' \cdot \alpha, u)$. The rest of the proof is similar to the above case.

∎